\documentclass[journal]{IEEEtran}
\usepackage{amsmath,amsthm,amssymb}
\usepackage[colorlinks=true]{hyperref}
\usepackage{algorithm}
\usepackage{algpseudocode}
\usepackage{graphicx}
\usepackage{cite}

\newtheorem{lemma}{Lemma}

\begin{document}
\title{PRIME: Phase Retrieval via Majorization-Minimization}
\author{Tianyu~Qiu, Prabhu~Babu, and~Daniel~P.~Palomar,~\IEEEmembership{Fellow,~IEEE}
\thanks{Tianyu Qiu, Prabhu Babu, and Daniel P. Palomar are with the Department of Electronic and Computer Engineering, The Hong Kong University of Science and Technology (HKUST), Hong Kong. Email:\{tqiu, eeprabhubabu, palomar\}@ust.hk.}
}



\maketitle

\begin{abstract}
This paper considers the phase retrieval problem in which measurements consist of only the magnitude of several linear measurements of the unknown, e.g., spectral components of a time sequence. We develop low-complexity algorithms with superior performance based on the majorization-minimization (MM) framework. The proposed algorithms are referred to as PRIME: Phase Retrieval vIa the Majorization-minimization techniquE. They are preferred to existing benchmark methods since at each iteration a simple surrogate problem is solved with a closed-form solution that monotonically decreases the original objective function. In total, four algorithms are proposed using different majorization-minimization techniques. Experimental results validate that our algorithms outperform existing methods in terms of successful recovery and mean square error under various settings.
\end{abstract}

\section{Introduction}
\IEEEPARstart{P}{hase} retrieval, the recovery of a signal from the magnitude of linear measurements like its Fourier transform, arises in various applications such as optical imaging \cite{Walther1963}, crystallography \cite{Harrison1993}, and microscopy \cite{Miao2002}. In general, optical devices (e.g., CCD cameras, human eyes, etc.) can record the intensity of the incoming light but not the phase, hence the problem of uniquely recovering the original signal is ill-posed due to the loss of phase information.

Mathematically speaking, the phase retrieval problem is to recover a $K$-dimensional complex signal $\mathbf{x}\in\mathbb{C}^K$ from the magnitude of $N$ linear measurements (usually corrupted with noise):
\begin{equation}
	y_i=\left|\mathbf{a}_i^H\mathbf{x}\right|^2+n_i\in\mathbb{R},\;i=1,\ldots,N,
\end{equation}
where the measurement vectors $\{\mathbf{a}_i\in\mathbb{C}^K\}_{i=1}^N$ are known beforehand. In the Fourier transform case, they correspond to rows of the Discrete Fourier Transform (DFT) matrix. In a more general case, they can be any vectors of interest. Due to the loss of phase information, the number of measurements should exceed the dimension of the original signal in order to successfully recover the signal. The authors of \cite{Candes2013b} proved that the number of measurements $N$ should at least be on the order of $K\log K$ for a successful recovery with high probability when the measurement vectors are chosen independently and uniformly at random on the unit sphere. A conjecture is posed in \cite{Bandeira2014} that $4K-4$ measurements are necessary and sufficient for injectivity, i.e., to uniquely recover the original signal (up to a constant phase shift) when provided with multiple measurements.

Numerical methods to recover the original signal $\mathbf{x}$ from multiple measurements $\{y_i\}_{i=1}^N$ fall mainly into two categories. The first is based on the Gerchberg-Saxton algorithm \cite{Gerchberg1972,Fienup1978,Netrapalli2013,Netrapalli2015}, and solves the phase retrieval problem through alternating minimizations. The second and more recent class is based on semidefinite relaxation \cite{Candes2013b,Candes2013a,Shechtman2015}. The idea is to recover the original signal through a convex semidefinite programming (SDP) problem by introducing a rank-$1$ matrix $\mathbf{X}:=\mathbf{xx}^H$, named ``matrix-lifting''. Unfortunately, the increase of dimension in this matrix-lifting procedure limits the application of the algorithm to small scale problems and it is not appropriate for big data problems. More recently, \cite{Candes2014a} proposed to solve the phase retrieval problem using the steepest descent method with a heuristic step size. Interestingly, one of the algorithms we present in this paper turns out to have similar updating rules but with a clearly specified step size. Besides these, other methods further exploit the signal sparsity: \cite{Shechtman2014} combined the damped Gauss-Newton method and ``2-opt" method to retrieve the phase of a sparse signal and \cite{Schniter2015} employed a probabilistic approach based on the generalized approximate message passing.

In this paper, we propose methods to solve the phase retrieval problem using different majorization-minimization (MM) techniques \cite{Hunter2004}. Instead of dealing with the original cumbersome optimization problem directly, an MM algorithm optimizes a sequence of simple surrogate problems successively. The sequence of points generated by solving each surrogate problem is guaranteed to converge to a stationary point of the original problem. As for the phase retrieval problem, by majorizing certain terms or all the terms in the objective function, we manage to substitute the original non-convex and difficult problem with different convex optimization problems. All these surrogate problems are designed to share the same favorable property of having a simple closed-form solution and only require basic matrix multiplications at every iteration. Different from the SDP approach, our algorithms do not require matrix-lifting, and at every iteration yield a simple closed-form solution directly for the signal $\mathbf{x}$. Therefore our algorithms can be applied to very large scale problems in big data systems.

The contributions of this paper are:
\begin{enumerate}
	\item Numerical methods for two different objectives of the phase retrieval problem (to recover the original signal from either the modulus squared or modulus of its linear measurements; problem \eqref{eq2} and problem \eqref{eq5}, respectively).
	\item Monotonicity and provable convergence to a stationary point of the sequence of points generated by our algorithms.
	\item Much faster numerical convergence of our algorithms (roughly four times faster according to simulation results).
	\item Low complexity per iteration of our algorithms (only requiring basic matrix multiplications).
\end{enumerate}

The remaining sections are organized as follows. In Section II, we present two different problem formulations for the phase retrieval problem. In Section III, after a brief overview of the general MM framework is introduced, we propose algorithms for both problems via different majorization-minimization techniques. An acceleration scheme is discussed in Section IV to further increase the convergence speed of our algorithms. Finally, in Section V, we provide the numerical results under various settings, e.g., different measurement matrices, clean measurements and noisy measurements.

\textit{Notation:} Boldface upper case letters (e.g., $\mathbf{X,A}$) denote matrices, while boldface lower case letters (e.g., $\mathbf{x,a}$) denote column vectors, and italics (e.g., $x,a,D$) denote scalars. $\mathbb{R}$ and $\mathbb{C}$ denote the real field and the complex field, respectively. For any complex number $x$, $|x|$ denotes the magnitude, and $\arg(x)$ denotes the phase. As for vectors, $|\mathbf{x}|$ denotes the element-wise magnitude and $\arg(\mathbf{x})$ denotes the element-wise phase. The superscripts $(\cdot)^T$, $\overline{(\cdot)}$ and $(\cdot)^H$ denote the transpose, complex conjugate and conjugate transpose, respectively. $X_{ij}$ or $[\mathbf{X}]_{ij}$ denotes the element at the $i$-th row and $j$-th column of a matrix $\mathbf{X}$, and $x_i$ or $[\mathbf{x}]_i$ denotes the $i$-th element of a vector $\mathbf{x}$. $\mathrm{Tr}(\cdot)$ is the trace of a matrix. $\mathrm{Diag}(\mathbf{x})$ is a diagonal matrix formed by setting vector $\mathbf{x}$ as its principal diagonal, while $\mathrm{diag}(\mathbf{X})$ is a column vector consisting of all the elements in the principal diagonal of matrix $\mathbf{X}$. The column vector $\mathrm{vec}(\mathbf{X})$ is formed by stacking all the columns of a matrix $\mathbf{X}$. As usual, the Euclidean norm of a vector $\mathbf{x}$ is denoted by $\|\mathbf{x}\|$. The curled inequality symbol $\succeq$ (its reverse form $\preceq$) is used to denote generalized inequality; $\mathbf{A}\succeq\mathbf{B}$ ($\mathbf{B}\preceq\mathbf{A}$) means that $\mathbf{A}-\mathbf{B}$ is a Hermitian positive semidefinite matrix. $\mathbf{I}_n$ is the $n\times n$ identity matrix, and $\mathbf{1}$ is a vector with all elements one.

\section{Problem Formulation and Existing Methods}
As described previously, the phase retrieval problem is to recover a complex signal from magnitudes of its linear measurements. In general, the problem is ill-posed due to the missing phase information. In this paper, we consider the case in which we have multiple measurements. Usually these measurements $\{y_i\}_{i=1}^N$ are corrupted with noise. When the noise follows a Gaussian distribution, a general choice is to consider the following least squares problem, which coincides with the maximum likelihood estimation of the original signal \cite{Shechtman2015,Candes2014a}:
\begin{equation}
	\label{eq2}
	\underset{\mathbf{x}}{\text{minimize}}\quad f(\mathbf{x}):=\sum\limits_{i=1}^{N}\left|y_i-\left|\mathbf{a}_i^H\mathbf{x}\right|^2\right|^2.
\end{equation}
Here, the measurement vectors $\{\mathbf{a}_i\in\mathbb{C}^K\}_{i=1}^N$ are known beforehand and can be any vectors of interest. In this paper, we consider two different cases. The first is the traditional Fourier transform case, in which $\{\mathbf{a}_i\}_{i=1}^N$ correspond to rows of the DFT matrix, i.e., the $k$-th element in vector $\mathbf{a}_i$ is $[\mathbf{a}_i]_k=e^{j2\pi(k-1)(i-1)/N}$. The second is the random matrix case, in which $\{\mathbf{a}_i\}_{i=1}^N$ are regarded as standard complex Gaussian distributed. Specifically, every element in the measurement vectors is a random variable in which both the real and imaginary parts are drawn from the standard Gaussian distribution $\mathcal{N}(0,1)$ independently.

Defining the measurement matrix
\begin{equation}
	\mathbf{A}=[\mathbf{a}_1,\mathbf{a}_2,\ldots,\mathbf{a}_N]\in\mathbb{C}^{K\times N},
\end{equation}
and stacking the multiple measurements $\{y_i\}_{i=1}^N$ as a vector $\mathbf{y}$, we can formulate the phase retrieval problem \eqref{eq2} in a more compact form:
\begin{equation}
	\underset{\mathbf{x}}{\text{minimize}}\quad\left\|\mathbf{y}-\left|\mathbf{A}^H\mathbf{x}\right|^2\right\|_2^2.
\end{equation}
Notice that here the operator $|\cdot|$ is applied element-wise when the argument is a vector (similarly for $(\cdot)^2$).

The authors of \cite{Candes2014a} proposed the following Wirtinger Flow algorithm based on the gradient descent method to solve problem \eqref{eq2}. They chose the leading eigenvector of $\sum_{i=1}^{N}y_i\mathbf{a}_i\mathbf{a}_i^H$ as the initial point because it would coincide with the optimal solution provided infinite samples ($N\rightarrow+\infty$) by the strong law of large samples.
\begin{algorithm}[H]
	\renewcommand{\algorithmicrequire}{\textbf{Input:}}
	\renewcommand{\algorithmicensure}{\textbf{Output:}}
	\caption{The Wirtinger Flow Algorithm}
	\begin{algorithmic}[1]
		\Require $\mathbf{A},\mathbf{y},t_0$
		\State Initial $\mathbf{x}^{(0)}\leftarrow$ leading eigenvector of $\sum\limits_{i=1}^{N}y_i\mathbf{a}_i\mathbf{a}_i^H$
		\State $\lambda^2=K\sum\limits_{i=1}^{N}y_i/\sum\limits_{i=1}^{N}\|\mathbf{a}_i\|^2$
		\State $\mathbf{x}^{(0)}\leftarrow\lambda\mathbf{x}^{(0)}$
		\For {$k=0,\ldots,t_0-1$}
		\State $\nabla f=4\sum\limits_{i=1}^{N}(|\mathbf{a}_i^H\mathbf{x}^{(k)}|^2-y_i)\mathbf{a}_i^H\mathbf{a}_i\mathbf{x}^{(k)}$
		\State $\mu^{(k+1)}=\lambda^2\min(1-e^{-\frac{k+1}{330}},0.4)$
		\State $\mathbf{x}^{(k+1)}=\mathbf{x}^{(k)}-\mu^{(k+1)}\nabla f$
		\EndFor
		\Ensure $\mathbf{x}^{(t_0)}$.
	\end{algorithmic}
\end{algorithm}

Different from problem \eqref{eq2}, an alternative is to solve the following problem using the modulus, as opposed to the squared modulus, of the linear measurements of the signal \cite{Gerchberg1972,Fienup1978,Netrapalli2013,Netrapalli2015}:
\begin{equation}
	\label{eq5}
	\underset{\mathbf{x}}{\text{minimize}}\quad\left\|\sqrt{\mathbf{y}}-\left|\mathbf{A}^H\mathbf{x}\right|\right\|_2^2,
\end{equation}
where the operator $\sqrt{\cdot}$ is applied element-wise when the argument is a vector. As pointed out in \cite{Netrapalli2013,Netrapalli2015}, if we had access to the phase information $\mathbf{c}$ of the linear measurements $\mathbf{A}^H\mathbf{x}$ (i.e., $c_i=e^{j\arg(\mathbf{a}_i^H\mathbf{x})}$) and $N\geq K$, then problem \eqref{eq5} would reduce to one of solving a system of linear equations
\begin{equation}
	\mathbf{C}\sqrt{\mathbf{y}}=\mathbf{A}^H\mathbf{x},
\end{equation}
where $\mathbf{C}:=\mathrm{Diag}(\mathbf{c})$ is a diagonal matrix. Of course we do not know this phase matrix $\mathbf{C}$; hence one intuitive approach is to solve the following problem by introducing a new variable $\mathbf{C}$ representing the phase information:
\begin{equation}
	\label{eq7}
	\begin{aligned}
		&\underset{\mathbf{x,C}}{\text{minimize}}&&\left\|\mathbf{A}^H\mathbf{x}-\mathbf{C}\sqrt{\mathbf{y}}\right\|_2^2\\
		&\text{subject to}&&\left|[\mathbf{C}]_{ii}\right|=1,\;i=1,\ldots,N,\\
		&&&[\mathbf{C}]_{ij}=0,\;i\neq j.
	\end{aligned}
\end{equation}
Note that the above problem \eqref{eq7} is not convex because the matrix $\mathbf{C}$ is restricted to be a diagonal matrix of phases; i.e., all the elements in the principal diagonal are limited to be of magnitude one, and zero elsewhere. One classical approach is to use the Gerchberg-Saxton algorithm \cite{Gerchberg1972}, thus alternately updating $\mathbf{x}$ and $\mathbf{C}$ so as to minimize problem \eqref{eq7}. For a given $\mathbf{C}$, problem \eqref{eq7} reduces to a standard least square problem, which can be solved easily. For a fixed $\mathbf{x}$, the optimal solution for $\mathbf{C}$ is $\mathbf{C}^\star=\mathrm{Diag}(e^{j\arg(\mathbf{A}^H\mathbf{x})})$. Here both the operators $e^{(\cdot)}$ and $\arg(\cdot)$ are applied element-wise.
\begin{algorithm}[H]
	\renewcommand{\algorithmicrequire}{\textbf{Input:}}
	\renewcommand{\algorithmicensure}{\textbf{Output:}}
	\caption{The Gerchberg-Saxton Algorithm}
	\label{alg2}
	\begin{algorithmic}[1]
		\Require $\mathbf{A},\mathbf{y},t_0$
		\State Initial $\mathbf{x}^{(0)}\leftarrow$ leading eigenvector of $\sum\limits_{i=1}^{N}y_i\mathbf{a}_i\mathbf{a}_i^H$
		\For {$k=0,\ldots,t_0-1$}
		\State $\mathbf{C}^{(k+1)}=\mathrm{Diag}\left(e^{j\arg\left(\mathbf{A}^H\mathbf{x}^{(k)}\right)}\right)$ 
		\State $\mathbf{x}^{(k+1)}=(\mathbf{AA}^H)^{-1}\mathbf{AC}^{(k+1)}\sqrt{\mathbf{y}}$
		\EndFor
		\Ensure $\mathbf{x}^{(t_0)}$.
	\end{algorithmic}
\end{algorithm}

In the next section, we are going to develop four algorithms using different MM techniques for both problems \eqref{eq2} and \eqref{eq5}. Experimental results show that our algorithms outperform the benchmark algorithms (Wirtinger Flow algorithm and Gerchberg-Saxton algorithm) in terms of successful recovery probability and mean square error.

\section{Phase Retrieval via Majorization-Minimization}
In this section, we first provide a concise introduction on the general MM framework, after which we present our algorithms for problems \eqref{eq2} and \eqref{eq5}. In total, four different algorithms are proposed, two for each problem. For problem \eqref{eq5}, one of the algorithms turns out to be exactly the same as the Gerchberg-Saxton algorithm. The other algorithm turns out to have similar updating rules to those of the Wirtinger Flow algorithm, but unlike in\cite{Candes2014a}, where a heuristic step size is used, our algorithm has a clearly specified step size. For problem \eqref{eq2}, one of the algorithms formulates it as the leading eigenvector problem, while the other algorithm further eliminates the requirement of the largest eigenvalue and corresponding eigenvector after one more majorization step.

\subsection{The MM Algorithm}
The majorization-minimization (MM) algorithm \cite{Hunter2004} is a generalization of the well-known expectation-maximization (EM) algorithm. Instead of dealing with the original difficult optimization problem directly, an MM algorithm solves a series of simple surrogate optimization problems, producing a series of points that drive the original objective function downhill.

For a real valued function $f(\boldsymbol{\theta})$, any function $g(\boldsymbol{\theta}\mid\boldsymbol{\theta}^{(m)})$ that satisfies the following two conditions is said to be a majorization function of $f(\boldsymbol{\theta})$ at the point $\boldsymbol{\theta}^{(m)}$:
\begin{equation}
	\label{eq8}
	\begin{aligned}
		g(\boldsymbol{\theta}\mid\boldsymbol{\theta}^{(m)})&\geq f(\boldsymbol{\theta})\text{ for all }\boldsymbol{\theta},\\
		g(\boldsymbol{\theta}^{(m)}\mid\boldsymbol{\theta}^{(m)})&=f(\boldsymbol{\theta}^{(m)}).
	\end{aligned}
\end{equation}
That is to say, the function $g(\boldsymbol{\theta}\mid\boldsymbol{\theta}^{(m)})$ is a global upper bound of the function $f(\boldsymbol{\theta})$, and touches it at the point $\boldsymbol{\theta}^{(m)}$. Instead of dealing with the original function $f(\boldsymbol{\theta})$ directly, which is usually non-convex or non-differentiable, an MM algorithm optimizes the sequence of majorization functions $\{g(\boldsymbol{\theta}\mid\boldsymbol{\theta}^{(m)})\}$. In general, these majorization functions are chosen to be convex and differentiable and much easier to solve, e.g., yielding a simple closed-form solution. Initialized by any feasible point $\boldsymbol{\theta}^{(0)}$, a sequence of points $\{\boldsymbol{\theta}^{(m)}\}$ is generated by the MM algorithm following the update rule:
\begin{equation}
	\label{eq9}
	\boldsymbol{\theta}^{(m+1)}\in\underset{\boldsymbol{\theta}}{\arg}\min g(\boldsymbol{\theta}\mid\boldsymbol{\theta}^{(m)}).
\end{equation}

A favorable property of the MM algorithm is that the sequence of points $\{\boldsymbol{\theta}^{(m)}\}$ generated by minimizing the majorization functions $\{g(\boldsymbol{\theta}\mid\boldsymbol{\theta}^{(m)})\}$ drive $f(\boldsymbol{\theta})$ downhill:
\begin{equation}
	\label{eq10}
	f(\boldsymbol{\theta}^{(m+1)})\leq g(\boldsymbol{\theta}^{(m+1)}\mid\boldsymbol{\theta}^{(m)})\leq g(\boldsymbol{\theta}^{(m)}\mid\boldsymbol{\theta}^{(m)})=f(\boldsymbol{\theta}^{(m)}).
\end{equation}
The first inequality and the third equality are a direct application of the definition of the majorization function in \eqref{eq8}. The second inequality comes from \eqref{eq9} that $\boldsymbol{\theta}^{(m+1)}$ is a minimizer of $g(\boldsymbol{\theta}\mid\boldsymbol{\theta}^{(m)})$. Hence under the MM framework, one can find a stationary point for the original function by solving the majorization functions instead.

\subsection{PRIME-Modulus-Single-Term}
We first apply the MM techniques to problem \eqref{eq5}. By expanding the objective function
\begin{equation}
	\left\|\sqrt{\mathbf{y}}-|\mathbf{A}^H\mathbf{x}|\right\|_2^2=\sum_{i=1}^N\left(\left|\mathbf{a}_i^H\mathbf{x}\right|^2-2\sqrt{y_i}\left|\mathbf{a}_i^H\mathbf{x}\right|+y_i\right)
\end{equation}
and discarding the constant term $\sum_{i=1}^Ny_i$, problem \eqref{eq5} is equivalent to
\begin{equation}
	\label{eq12}
	\underset{\mathbf{x}}{\text{minimize}}\quad\sum\limits_{i=1}^{N}\left(\left|\mathbf{a}_i^H\mathbf{x}\right|^2-2\sqrt{y_i}\left|\mathbf{a}_i^H\mathbf{x}\right|\right).
\end{equation}

Here we keep the first term $\sum_{i=1}^N\left|\mathbf{a}_i^H\mathbf{x}\right|^2$ and only majorize the (nonconvex) second term $-\sum_{i=1}^N2\sqrt{y_i}\left|\mathbf{a}_i^H\mathbf{x}\right|$. According to the Cauchy-Schwarz inequality
\begin{equation}
	\left|\mathbf{a}_i^H\mathbf{x}\right|\cdot\left|\mathbf{a}_i^H\mathbf{x}^{(k)}\right|\geq\mathrm{Re}\left(\mathbf{a}_i^H\mathbf{x}\cdot(\mathbf{x}^{(k)})^H\mathbf{a}_i\right),
\end{equation}
the second term $-\sum_{i=1}^N2\sqrt{y_i}\left|\mathbf{a}_i^H\mathbf{x}\right|$ can be majorized as
\begin{equation}
	-\sum\limits_{i=1}^N2\sqrt{y_i}\left|\mathbf{a}_i^H\mathbf{x}\right|\leq-\sum\limits_{i=1}^N2\sqrt{y_i}\frac{\mathrm{Re}\left(\mathbf{a}_i^H\mathbf{x}\cdot(\mathbf{x}^{(k)})^H\mathbf{a}_i\right)}{\left|\mathbf{a}_i^H\mathbf{x}^{(k)}\right|}.
\end{equation}
Thus the convex majorization problem for \eqref{eq12} is 
\begin{equation}
	\underset{\mathbf{x}}{\text{minimize}}\quad\sum\limits_{i=1}^{N}\left(\left|\mathbf{a}_i^H\mathbf{x}\right|^2-2\sqrt{y_i}\frac{\mathrm{Re}\left(\mathbf{a}_i^H\mathbf{x}\cdot(\mathbf{x}^{(k)})^H\mathbf{a}_i\right)}{\left|\mathbf{a}_i^H\mathbf{x}^{(k)}\right|}\right),
\end{equation}
which is equivalent to
\begin{equation}
	\label{eq16}
	\underset{\mathbf{x}}{\text{minimize}}\quad\sum\limits_{i=1}^{N}\left|\mathbf{a}_i^H\mathbf{x}-c_i\right|^2,
\end{equation}
where
\begin{equation}
	c_i:=\frac{\sqrt{y_i}\mathbf{a}_i^H\mathbf{x}^{(k)}}{\left|\mathbf{a}_i^H\mathbf{x}^{(k)}\right|}.
\end{equation}
Note that $c_i$ actually stands for the phase information of the linear measurement $\mathbf{a}_i^H\mathbf{x}^{(k)}$, only scaled by a constant value $\sqrt{y_i}$. Therefore, we introduce here the matrix $\mathbf{C}:=\mathrm{Diag}(e^{j\arg\left(\mathbf{A}^H\mathbf{x}^{(k)}\right)})$ and further can formulate problem \eqref{eq16} in the following more compact form:
\begin{equation}
	\label{eq18}
	\underset{\mathbf{x}}{\text{minimize}}\quad\left\|\mathbf{A}^H\mathbf{x}-\mathbf{C}\sqrt{\mathbf{y}}\right\|_2^2,
\end{equation}
which is a simple least square problem. And it has a simple closed-form solution $\mathbf{x}^{\star}=(\mathbf{AA}^H)^{-1}\mathbf{AC}\sqrt{\mathbf{y}}$ if the measurement matrix $\mathbf{A}$ has full row rank. It is quite interesting that we have managed to solve problem \eqref{eq5} deriving the same Gerchberg-Saxton algorithm but from a totally different majorization-minimization perspective.

\subsection{PRIME-Modulus-Both-Terms}
We now consider majorizing both terms in problem \eqref{eq12}. In principle, this is not necessary since after majorizing the second term the problem becomes convex with a simple closed-form solution. Also, in general, the fewer terms we majorize, the better it tends to be in terms of convergence. Nevertheless, we explore this option for the sake of an even simpler algorithm, albeit with a potentially slower convergence.

\begin{lemma}\cite{Song2014}
	\label{lemma1}
	Let $\mathbf{L}$ be a $K\times K$ Hermitian matrix and $\mathbf{M}$ be another $K\times K$ Hermitian matrix such that $\mathbf{M}\succeq\mathbf{L}$. Then for any point $\mathbf{x}_0\in\mathbb{C}^K$, the quadratic function $\mathbf{x}^H\mathbf{Lx}$ is majorized by $\mathbf{x}^H\mathbf{Mx}+2\mathrm{Re}\left(\mathbf{x}^H\left(\mathbf{L}-\mathbf{M}\right)\mathbf{x}_0\right)+\mathbf{x}_0^H\left(\mathbf{M}-\mathbf{L}\right)\mathbf{x}_0$ at $\mathbf{x}_0$.
\end{lemma}
The above lemma provides a method to majorize the first term $\sum_{i=1}^{N}\left|\mathbf{a}_i^H\mathbf{x}\right|^2$.
\begin{equation}
	\begin{aligned}
		&\sum\limits_{i=1}^{N}\left|\mathbf{a}_i^H\mathbf{x}\right|^2=\sum\limits_{i=1}^{N}\mathbf{x}^H\mathbf{a}_i\mathbf{a}_i^H\mathbf{x}=\mathbf{x}^H\mathbf{AA}^H\mathbf{x}\\
		\leq&\lambda_{\max}(\mathbf{AA}^H)\mathbf{x}^H\mathbf{x}\\
		&+ 2\mathrm{Re}\left[\mathbf{x}^H\left(\mathbf{AA}^H-\lambda_{\max}(\mathbf{AA}^H)\mathbf{I}\right)\mathbf{x}^{(k)}\right]\\
		&+(\mathbf{x}^{(k)})^H\left(\lambda_{\max}(\mathbf{AA}^H)\mathbf{I}-\mathbf{AA}^H\right)\mathbf{x}^{(k)}.
	\end{aligned}
\end{equation}
Discarding the constant term, the new majorization problem for \eqref{eq12} is
\begin{equation}
	\label{eq20}
	\begin{aligned}
		&\underset{\mathbf{x}}{\text{minimize}}\quad\lambda_{\max}(\mathbf{AA}^H)\mathbf{x}^H\mathbf{x}\\
		+&2\mathrm{Re}\left[\mathbf{x}^H\left(\mathbf{AA}^H-\lambda_{\max}(\mathbf{AA}^H)\mathbf{I}-\sum\limits_{i=1}^{N}\frac{\sqrt{y_i}\mathbf{a}_i\mathbf{a}_i^H}{\left|\mathbf{a}_i^H\mathbf{x}^{(k)}\right|}\right)\mathbf{x}^{(k)}\right],
	\end{aligned}
\end{equation}
which is equivalent to the following problem:
\begin{equation}
	\label{eq21}
	\underset{\mathbf{x}}{\text{minimize}}\quad\left\|\mathbf{x}-\mathbf{c}\right\|_2^2,
\end{equation}
and has a closed-form solution $\mathbf{x}^\star=\mathbf{c}$, where the constant
\begin{equation}
	\mathbf{c}:=\mathbf{x}^{(k)}+\lambda_{\max}^{-1}(\mathbf{AA}^H)\left(\sum\limits_{i=1}^{N}\frac{\sqrt{y_i}\mathbf{a}_i\mathbf{a}_i^H}{\left|\mathbf{a}_i^H\mathbf{x}^{(k)}\right|}-\mathbf{AA}^H\right)\mathbf{x}^{(k)}.
\end{equation}
This algorithm is similar to the steepest descent method proposed recently in \cite{Candes2014a}, where the authors chose a heuristic step size. Now we can see that one suitable step size is $\lambda_{\max}^{-1}(\mathbf{AA}^H)$.

So far there is no preference between these two majorization problems for problem \eqref{eq5}. They both yield a simple closed-form solution at every iteration and are preferable under different problem settings. The procedure to solve the first majorization problem \eqref{eq18} turns out to be the same as in the Gerchberg-Saxton algorithm, in which at every iteration one only needs to solve a standard least square problem. As for the second majorization problem \eqref{eq21}, one needs to calculate the leading eigenvalue $\lambda_{\max}(\mathbf{AA}^H)$, which is cumbersome when the signal to be recovered is of a very large dimension. Fortunately in the Fourier transform case, when the measurement matrix $\mathbf{A}$ is from the DFT matrix, this largest eigenvalue is as simple as $\lambda_{\max}(\mathbf{AA}^H)=N$ (see Appendix A for the proof).

\subsection{PRIME-Power}
Now we are going to show step by step how we majorize problem \eqref{eq2} as a leading eigenvector problem using MM techniques. By introducing two matrices $\mathbf{A}_i=\mathbf{a}_i\mathbf{a}_i^H$ and $\mathbf{X}=\mathbf{xx}^H$, we can rewrite problem \eqref{eq2} as
\begin{equation}
	\begin{aligned}
		& \underset{\mathbf{x,X}}{\text{minimize}}&&\sum\limits_{i=1}^{N}\left(y_i-\mathrm{Tr}(\mathbf{A}_i\mathbf{X})\right)^2 \\
		& \text{subject to} && \mathbf{X}=\mathbf{xx}^H,
	\end{aligned}
\end{equation}
which is equivalent to 
\begin{equation}
	\label{eq24}
	\begin{aligned}
		&\underset{\mathbf{x,X}}{\text{minimize}}&&\sum\limits_{i=1}^{N}\left[\left(\mathrm{Tr}(\mathbf{A}_i\mathbf{X})\right)^2-2y_i\mathrm{Tr}(\mathbf{A}_i\mathbf{X})\right] \\
		&\text{subject to} && \mathbf{X}=\mathbf{xx}^H,
	\end{aligned}
\end{equation}
by ignoring the constant term $\sum_{i=1}^Ny_i^2$. We choose to majorize the first term $\sum_{i=1}^N\left(\mathrm{Tr}(\mathbf{A}_i\mathbf{X})\right)^2$ (note that this term is already convex in $\mathbf{X}$ but the majorization will help in producing a much simpler problem), and keep the second term $\sum_{i=1}^N(-2y_i\mathrm{Tr}(\mathbf{A}_i\mathbf{X}))$ since it is linear in $\mathbf{X}$.

Note that both matrices $\mathbf{X}$ and $\mathbf{A}_i$ are Hermitian. Thus we can write the first term $\sum_{i=1}^N\left(\mathrm{Tr}(\mathbf{A}_i\mathbf{X})\right)^2$ in problem \eqref{eq24} as
\begin{equation}
	\label{eq25}
	\begin{aligned}
		\sum\limits_{i=1}^{N}\left(\mathrm{Tr}(\mathbf{A}_i\mathbf{X})\right)^2&=\sum\limits_{i=1}^{N}\mathrm{vec}(\mathbf{X})^H\mathrm{vec}(\mathbf{A}_i)\mathrm{vec}(\mathbf{A}_i)^H\mathrm{vec}(\mathbf{X})\\
		&=\mathrm{vec}(\mathbf{X})^H\mathbf{\Phi}\mathrm{vec}(\mathbf{X}),\\
	\end{aligned}
\end{equation}
where we define the matrix
\begin{equation}
	\mathbf{\Phi}:=\sum_{i=1}^N\mathrm{vec}(\mathbf{A}_i)\mathrm{vec}(\mathbf{A}_i)^H.
\end{equation}
This matrix $\mathbf{\Phi}$ is just a constant with regard to the variables $\mathbf{x}$ and $\mathbf{X}$ since all the measurement vectors $\{\mathbf{a}_i\}_{i=1}^N$ are known beforehand. According to Lemma \ref{lemma1}, by treating the matrix $\mathbf{\Phi}$ as $\mathbf{L}$ and setting $\mathbf{M}=D\mathbf{I}_{K^2}$, where $D\geq\lambda_{\max}(\mathbf{\Phi})$ guarantees $\mathbf{M}\succeq\mathbf{L}$, the expression $\mathrm{vec}(\mathbf{X})^H\mathbf{\Phi}\mathrm{vec}(\mathbf{X})$ in \eqref{eq25} can be majorized by the following function (from now on and when no misunderstanding is caused, the dimension in the identity matrix will be omitted for the sake of notation):
\begin{equation}
	\label{eq27}
	\begin{aligned}
		& u_1(\mathbf{X}\mid\mathbf{X}^{(k)})\\
		=&D\mathrm{vec}(\mathbf{X})^H\mathrm{vec}(\mathbf{X})+2\mathrm{Re}\left[\mathrm{vec}(\mathbf{X})^H\left(\mathbf{\Phi}-D\mathbf{I}\right)\mathrm{vec}(\mathbf{X}^{(k)})\right]\\
		&+\mathrm{vec}(\mathbf{X}^{(k)})^H\left(D\mathbf{I}-\mathbf{\Phi}\right)\mathrm{vec}(\mathbf{X}^{(k)})\\
		=&D\mathrm{Tr}(\mathbf{XX})+2\sum\limits_{i=1}^{N}\mathrm{Tr}(\mathbf{XA}_i)\mathrm{Tr}(\mathbf{X}^{(k)}\mathbf{A}_i)-2D\mathrm{Tr}(\mathbf{XX}^{(k)})\\
		&+const.,
	\end{aligned}
\end{equation}
where $const.$ represents a constant term not dependent on $\mathbf{X}$. Combining this majorization function $u_1(\mathbf{X}\mid\mathbf{X}^{(k)})$ and the second term $\sum_{i=1}^N(-2y_i\mathrm{Tr}(\mathbf{A}_i\mathbf{X}))$ in problem \eqref{eq24} together, and discarding the constant terms, we can get the majorization problem for \eqref{eq24} as
\begin{equation}
	\label{eq28}
	\begin{aligned}
		&\underset{\mathbf{x,X}}{\text{minimize}}&&D\mathrm{Tr}(\mathbf{XX})+2\sum\limits_{i=1}^{N}\mathrm{Tr}(\mathbf{XA}_i)\mathrm{Tr}(\mathbf{X}^{(k)}\mathbf{A}_i)\\
		&&&-2D\mathrm{Tr}(\mathbf{XX}^{(k)})-2\sum\limits_{i=1}^{N}y_i\mathrm{Tr}(\mathbf{A}_i\mathbf{X}) \\
		&\text{subject to} && \mathbf{X}=\mathbf{xx}^H,
	\end{aligned}
\end{equation}
which is equivalent to the following leading eigenvector problem:
\begin{equation}
	\label{eq29}
	\begin{aligned}
		&\underset{\mathbf{x,X}}{\text{minimize}}&&\|\mathbf{X-W}\|_F^2 \\
		&\text{subject to} && \mathbf{X}=\mathbf{xx}^H,\\
	\end{aligned}
\end{equation}
with the matrix
\begin{equation}
	\label{eq30}
	\mathbf{W}:=\mathbf{X}^{(k)}+\frac{1}{D}\sum\limits_{i=1}^N\left(y_i-\mathrm{Tr}(\mathbf{X}^{(k)}\mathbf{A}_i)\right)\mathbf{A}_i.
\end{equation}
The solution to this leading eigenvector problem is
\begin{equation}
	\left\{
	\begin{aligned}
		\mathbf{x}^\star&=\sqrt{\lambda_{\max}(\mathbf{W})}\mathbf{u}_{\max}(\mathbf{W}),\\
		\mathbf{X}^\star&=\mathbf{x}^\star(\mathbf{x}^\star)^H.
	\end{aligned}
	\right.
\end{equation}
where $\lambda_{\max}(\mathbf{W})$ and $\mathbf{u}_{\max}(\mathbf{W})$ are the largest eigenvalue and corresponding eigenvector of matrix $\mathbf{W}$. The procedures are summarized in Algorithm \ref{alg3}. A general choice is to conduct eigen-decomposition to find this largest eigenvalue and corresponding eigenvector, which unfortunately is usually computationally costly and time consuming. Therefore, we propose to use the power iteration method instead without conducting the eigen-decomposition. The power iteration method is a simple iterative algorithm to calculate the eigenvalue (the one with the greatest absolute value) and corresponding eigenvector. Together with the following lemma, the power iteration method will indeed produce the largest eigenvalue and the corresponding eigenvector.
\begin{lemma}
	\label{lemma2}
	For the matrix $\mathbf{W}$ defined in \eqref{eq30}, its largest eigenvalue and smallest eigenvalue satisfy the inequality
	\begin{equation}
		\lambda_{\max}(\mathbf{W})>|\lambda_{\min}(\mathbf{W})|,
	\end{equation}
	provided that
	\begin{equation}
		\begin{aligned}
			D>&\sum\limits_{i\in\mathcal{I}}\left(\left|\mathbf{a}_i^H\mathbf{x}^{(k)}\right|^2-y_i\right)\frac{\|\mathbf{a}_i\|^2}{\|\mathbf{x}^{(k)}\|^2}\\
			&+\sum\limits_{i=1}^{N}\left(\left|\mathbf{a}_i^H\mathbf{x}^{(k)}\right|^2-y_i\right)\frac{\left|\mathbf{a}_i^H\mathbf{x}^{(k)}\right|^2}{\|\mathbf{x}^{(k)}\|^4},
		\end{aligned}
	\end{equation}
	where the set $\mathcal{I}:=\{i:y_i<\left|\mathbf{a}_i^H\mathbf{x}^{(k)}\right|^2\}$.
\end{lemma}
\begin{IEEEproof}
	Appendix B.
\end{IEEEproof}

\begin{algorithm}[H]
	\renewcommand{\algorithmicrequire}{\textbf{Input:}}
	\renewcommand{\algorithmicensure}{\textbf{Output:}}
	\caption{PRIME-Power}
	\label{alg3}
	\begin{algorithmic}[1]
		\Require $\mathbf{A},\mathbf{y},t_0$
		\State Initial $\mathbf{x}^{(0)}\leftarrow$ leading eigenvector of $\mathbf{B}:=\sum\limits_{i=1}^{N}y_i\mathbf{A}_i$
		\For {$k=0,\ldots,t_0-1$}
		\State $\mathbf{W}=\mathbf{x}^{(k)}(\mathbf{x}^{(k)})^H+\frac{1}{D}\left(\mathbf{B}-\sum\limits_{i=1}^{N}\left|\mathbf{a}_i^H\mathbf{x}^{(k)}\right|^2\mathbf{A}_i\right)$
		\State $\mathbf{x}^{(k+1)}=\sqrt{\lambda_{\max}(\mathbf{W})}\mathbf{u}_{\max}(\mathbf{W})$
		\EndFor
		\Ensure $\mathbf{x}^{(t_0)}$.
	\end{algorithmic}
\end{algorithm}

\subsection{PRIME-Power-Backtracking}
Instead of formulating problem \eqref{eq28} as the leading eigenvector problem \eqref{eq29}, which still takes some effort to find the largest eigenvalue and corresponding eigenvector, in this subsection we will show how to formulate it as a problem that does not require such information.

Since $\mathbf{X}=\mathbf{xx}^H$, we can write it as $\mathbf{X}=t\tilde{\mathbf{x}}\tilde{\mathbf{x}}^H$, where $t=\|\mathbf{x}\|^2$ and $\tilde{\mathbf{x}}=\mathbf{x}/\|\mathbf{x}\|$. Then problem \eqref{eq28} is equivalent to
\begin{equation}
	\label{eq34}
	\begin{aligned}
		&\underset{\tilde{\mathbf{x}},\;t\geq 0}{\text{minimize}}&&t^2-2t\tilde{\mathbf{x}}^H\mathbf{W}\tilde{\mathbf{x}}\\
		&\text{subject to} && \|\tilde{\mathbf{x}}\|=1,
	\end{aligned}
\end{equation}
with $\mathbf{W}$ defined in \eqref{eq30}. For a fixed $t\geq 0$, we need to solve the following problem:
\begin{equation}
	\label{eq35}
	\begin{aligned}
		&\underset{\tilde{\mathbf{x}}}{\text{minimize}}&&-\tilde{\mathbf{x}}^H\mathbf{W}\tilde{\mathbf{x}}\\
		&\text{subject to} && \|\tilde{\mathbf{x}}\|=1.
	\end{aligned}
\end{equation}
The solution is the normalized largest eigenvector of $\mathbf{W}$. Instead of using the power iteration method, we now apply a second majorization step to problem \eqref{eq35}. According to Lemma \ref{lemma1}, setting $-\mathbf{W}$ as $\mathbf{L}$ and $E\mathbf{I}$ as $\mathbf{M}$ ($E\geq\lambda_{\max}(-\mathbf{W})=\lambda_{\min}(\mathbf{W})$) guarantees $\mathbf{M}\succeq\mathbf{L}$. The corresponding majorization problem is
\begin{equation}
	\label{eq36}
	\begin{aligned}
		&\underset{\tilde{\mathbf{x}}}{\text{minimize}}&&-2\mathrm{Re}\left(\tilde{\mathbf{x}}^H\mathbf{d}\right)\\
		&\text{subject to} &&\|\tilde{\mathbf{x}}\|=1,\\
	\end{aligned}
\end{equation}
with the constant term $\mathbf{d}:=(\mathbf{W}+E\mathbf{I})\tilde{\mathbf{x}}^{(k)}$, and it has a closed-form solution $\tilde{\mathbf{x}}^{\star}=\mathbf{d}/\left\|\mathbf{d}\right\|$. On the other hand, for a fixed $\tilde{\mathbf{x}}$, we need to solve the following problem:
\begin{equation}
	\underset{\;t\geq 0}{\text{minimize}}\quad t^2-2t\tilde{\mathbf{x}}^H\mathbf{W}\tilde{\mathbf{x}}.
\end{equation}
It has a closed-form solution $t^{\star}=\max\{0,\;\tilde{\mathbf{x}}^H\mathbf{W}\tilde{\mathbf{x}}\}$. Therefore after applying the second majorization step, we get a simple closed-form solution at every iteration. We can also combine the two majorization functions together as
\begin{equation}
	\begin{aligned}
		&f(\mathbf{x})=\sum\limits_{i=1}^{N}\left|y_i-\left|\mathbf{a}_i^H\mathbf{x}\right|^2\right|^2\leq D\|\mathbf{x}\|^4\\		+&2D\|\mathbf{x}\|^2\tilde{\mathbf{x}}^H(-\mathbf{W})\tilde{\mathbf{x}}+D\|\mathbf{x}^{(k)}\|^4-\sum\limits_{i=1}^{N}\left|\mathbf{a}_i^H\mathbf{x}^{(k)}\right|^4+\sum\limits_{i=1}^{N}y_i^2\\
		\leq&D\|\mathbf{x}\|^4+2DE\|\mathbf{x}\|^2-4D\frac{\|\mathbf{x}\|}{\|\mathbf{x}^{(k)}\|}\mathrm{Re}\left[\mathbf{x}^H(\mathbf{W}+E\mathbf{I})\mathbf{x}^{(k)}\right]\\
		&+2D\frac{\|\mathbf{x}\|^2}{\|\mathbf{x}^{(k)}\|^2}(\mathbf{x}^{(k)})^H(\mathbf{W}+E\mathbf{I})\mathbf{x}^{(k)}+D\|\mathbf{x}^{(k)}\|^4\\
		&-\sum\limits_{i=1}^{N}\left|\mathbf{a}_i^H\mathbf{x}^{(k)}\right|^4+\sum\limits_{i=1}^{N}y_i^2\\
		:=&g(\mathbf{x}\mid\mathbf{x}^{(k)}).
	\end{aligned}
\end{equation}
Regarding the right choice of $E$, a simple backtracking method should be sufficient to ensure a valid majorization function. In other words, after the two majorization steps \eqref{eq28} and \eqref{eq36}, the descent property \eqref{eq10} holds. The procedures are summarized in Algorithm \ref{alg4}. (We choose $D=\lambda_{\max}(\mathbf{\Phi})$ here.)

\begin{algorithm}[H]
	\renewcommand{\algorithmicrequire}{\textbf{Input:}}
	\renewcommand{\algorithmicensure}{\textbf{Output:}}
	\caption{PRIME-Power-Backtracking}
	\label{alg4}
	\begin{algorithmic}[1]
		\Require $\mathbf{A},\mathbf{y},t_0$
		\State Initial $\mathbf{x}^{(0)}\leftarrow$ leading eigenvector of $\mathbf{B}:=\sum\limits_{i=1}^{N}y_i\mathbf{A}_i$
		\For {$k=0,\ldots,t_0-1$}
		\State $\mathbf{W}=\mathbf{x}^{(k)}(\mathbf{x}^{(k)})^H+\frac{1}{D}\left(\mathbf{B}-\sum\limits_{i=1}^{N}\left|\mathbf{a}_i^H\mathbf{x}^{(k)}\right|^2\mathbf{A}_i\right)$
		\State $\tilde{\mathbf{x}}^{(k)}=\mathbf{x}^{(k)}/\|\mathbf{x}^{(k)}\|$
		\State $E=0.5$
		\Repeat 
		\State $E\leftarrow 2E$
		\State $\mathbf{d}=(\mathbf{W}+E\mathbf{I})\tilde{\mathbf{x}}^{(k)}$
		\State $\tilde{\mathbf{x}}^{(k+1)}=\mathbf{d}/\left\|\mathbf{d}\right\|$
		\State $t^{(k+1)}=\max\{0,\;(\tilde{\mathbf{x}}^{(k+1)})^H\mathbf{W}\tilde{\mathbf{x}}^{(k+1)}\}$
		\State $\mathbf{x}^{(k+1)}=\sqrt{t^{(k+1)}}\tilde{\mathbf{x}}^{(k+1)}$
		\Until {$g(\mathbf{x}^{(k+1)}\mid\mathbf{x}^{(k)})\geq f(\mathbf{x}^{(k+1)})$}
		\EndFor
		\Ensure $\mathbf{x}^{(t_0)}$.
	\end{algorithmic}
\end{algorithm}

\subsection{Convergence Analysis}
Inherited from the general majorization-minimization framework, the non-increasing property \eqref{eq10} holds for any majorization problem. And the objective value is bounded below by $0$ either for problem \eqref{eq2} or for problem \eqref{eq5}. Therefore the sequence $\{f(\mathbf{x}^{(k)})\}$ generated by our algorithms is guaranteed to converge to a finite point at least. Moreover, the authors of \cite{Razaviyayn2013} proved that any sequence $\{\mathbf{x}^{(k)}\}$ generated by the MM algorithm converges to a stationary point when the constraint set is closed and convex. Unfortunately the two majorization problems \eqref{eq28} and \eqref{eq36} for problem \eqref{eq2} involve a non-convex constraint set. Therefore we prove below that the sequence $\{\mathbf{x}^{(k)}\}$ generated by \eqref{eq28} and \eqref{eq36} also converges to a stationary point.
\begin{lemma}
	The sequence $\{\mathbf{x}^{(k)}\}$ generated by majorization problem \eqref{eq28} (also \eqref{eq36}) converges to a stationary point of problem \eqref{eq2}.
\end{lemma}
\begin{IEEEproof}
	We have shown above that problem \eqref{eq28} is equivalent to \eqref{eq34} by a change of variables. And problem \eqref{eq34} has the same solution as the following problem:
	\begin{equation}
	\label{eq39}
	\begin{aligned}
	&\underset{\tilde{\mathbf{x}},\;t\geq 0}{\text{minimize}}&&t^2-2t\tilde{\mathbf{x}}^H\mathbf{W}\tilde{\mathbf{x}}\\
	&\text{subject to} && \|\tilde{\mathbf{x}}\|\leq 1.
	\end{aligned}
	\end{equation}
	Assume \eqref{eq39} has an optimal solution $\{t^{\star},\tilde{\mathbf{x}}^{\star}\}$ and $\|\tilde{\mathbf{x}}^{\star}\|<1$. Then $\{t^{\star},\tilde{\mathbf{x}}^{\star}/\|\tilde{\mathbf{x}}^{\star}\|\}$ is feasible and achieves a smaller objective value. This contradicts the assumption that $\{t^{\star},\tilde{\mathbf{x}}^{\star}\}$ is an optimal solution. Hence, any optimal solution $\{t^{\star},\tilde{\mathbf{x}}^{\star}\}$ for \eqref{eq39} should satisfy $\|\tilde{\mathbf{x}}^{\star}\|=1$. Now if we relax \eqref{eq34} to \eqref{eq39}, which has a closed convex constraint set, the sequence $\{\mathbf{x}^{k}\}$ generated by \eqref{eq39} should converge to a stationary point of problem \eqref{eq2} following from Theorem 1 in \cite{Razaviyayn2013}.
	
	Similarly, we can prove that the sequence $\{\mathbf{x}^{(k)}\}$ generated by \eqref{eq36} also converges to a stationary point of problem \eqref{eq2}.
\end{IEEEproof}

\subsection{Computational Complexity}
We now offer a short discussion on the computational complexity of all the algorithms we have proposed so far. The two algorithms for problem \eqref{eq5}, namely, PRIME-Modulus-Single-Term and PRIME-Modulus-Both-Terms, both yield a simple closed-form solution and only require basic matrix complications at every iteration. The time complexities for these two algorithms are $O(NK^2)$ and $O(NK)$, respectively. For problem \eqref{eq2}, PRIME-Power needs the leading eigenvalue and corresponding eigenvector of an intermediate matrix at every iteration, while PRIME-Power-Backtracking avoids this by introducing an inner loop. The time complexities of these two algorithms are both $O(NK^2)$.

\section{Acceleration Scheme}
The popularity of the MM framework is attributed to its simplicity and monotonic decreasing property, at the expense of a usually low convergence rate. This slow convergence may jeopardize the performance of the MM algorithm for computing intensive tasks. In \cite{VARADHAN2008}, the authors proposed a simple and globally convergent method to accelerate any EM algorithms. This accelerating algorithm, called the squared iterative methods (SQUAREM), generally achieves superlinear convergence, and is especially attractive in high-dimensional problems as it only requires parameter updating. Besides this, since the MM algorithm is a generalization of the EM algorithm, SQUAREM can be adopted to update the parameters in our MM-based algorithms. At every iteration, instead of updating $\mathbf{x}^{(k+1)}$ directly from $\mathbf{x}^{(k)}$, SQUAREM seeks an intermediate point $\mathbf{x}'$ from $\mathbf{x}^{(k)}$, after which, it updates the next point $\mathbf{x}^{(k+1)}$ based on this intermediate point.

\section{Numerical Simulations}
In this section, we present the experimental results for both problem \eqref{eq2} and problem \eqref{eq5} under various settings. Specifically, we consider that the measurement matrix is either standard complex Gaussian distributed or from the DFT matrix, and the measurements are clean or corrupted with Gaussian noise. All experiments are conducted on a personal computer with a 3.20 GHz Intel Core i5-4570 CPU and 8.00 GB RAM.

For both problems, our MM-based algorithms outperform the benchmark methods, Wirtinger Flow algorithm and Gerchberg-Saxton algorithm, respectively, in terms of successful recovery probability and convergence speed. Details of the experiments and comparisons can be found in later subsections under different settings.

\subsection{Random Gaussian Matrix Setting}
First we consider the case in which all the elements in the measurement matrix $\mathbf{A}$ are independent random variables following a standard complex Gaussian distribution. Thus every element is regarded as a random variable in which the real part and the imaginary part are drawn from the standard Gaussian distribution $\mathcal{N}(0,1)$ independently.

We choose a random signal $\mathbf{x}_o\in\mathbb{C}^{10}$ (normalized to $\mathbf{x}_o/\|\mathbf{x}_o\|$ without loss of generality) as the original signal, and generate the measurements $\mathbf{y}=|\mathbf{A}^H\mathbf{x}_o|^2\in\mathbb{R}^N$ accordingly. Since the measurement matrix $\mathbf{A}$ is a random matrix here, we repeat the experiments $1000$ times using different and independent measurement matrices, with everything else fixed as the same. In the PRIME-Power algorithm we propose to use the power iteration method to calculate the largest eigenvalue and corresponding eigenvector instead of conducting eigen-decomposition. Experimental results indicate that one step of the power iteration is sufficient enough to considerably reduce the computations without degrading the performance. As for the Wirtinger Flow algorithm, different from the heuristic step size used in the original paper \cite{Candes2014a}, here we adopt a  backtracking method to find a suitable step size. We also use the fixed point method to accelarate our algorithms, which leads to the names PRIME-Power-Acce, PRIME-Power-Backtracking-Acce, PRIME-Modulus-Single-Term-Acce, and PRIME-Modulus-Both-Terms-Acce accordingly.

As mentioned above, all the algorithms can only recover the original signal $\mathbf{x}_o$ up to a constant phase shift due to the loss of phase information. Fortunately, we can easily find this constant phase by the following procedure. For any solution $\mathbf{x}^{\star}$ returned from the algorithms, we define a function

\begin{figure}[t]
	\centering
	\includegraphics[width=9cm]{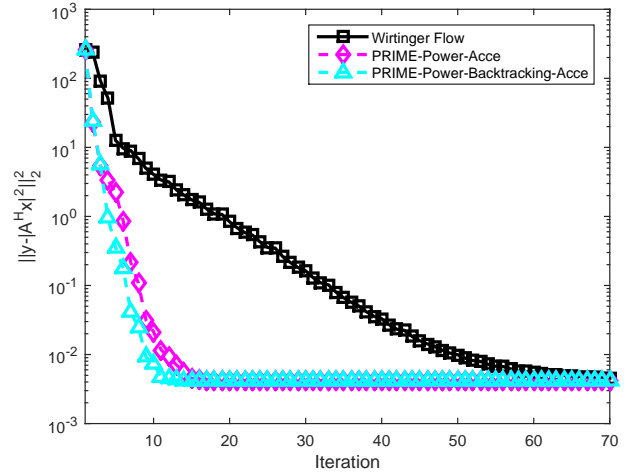}
	\caption{Objective function versus iteration for problem \eqref{eq2} under noisy measurements and random Gaussian matrix setting. $\mathbf{x}\in\mathbb{C}^{10},\mathbf{y}\in\mathbb{R}^{50}$.}
	\label{fig1}
\end{figure}
\begin{figure}[t]
	\centering
	\includegraphics[width=9cm]{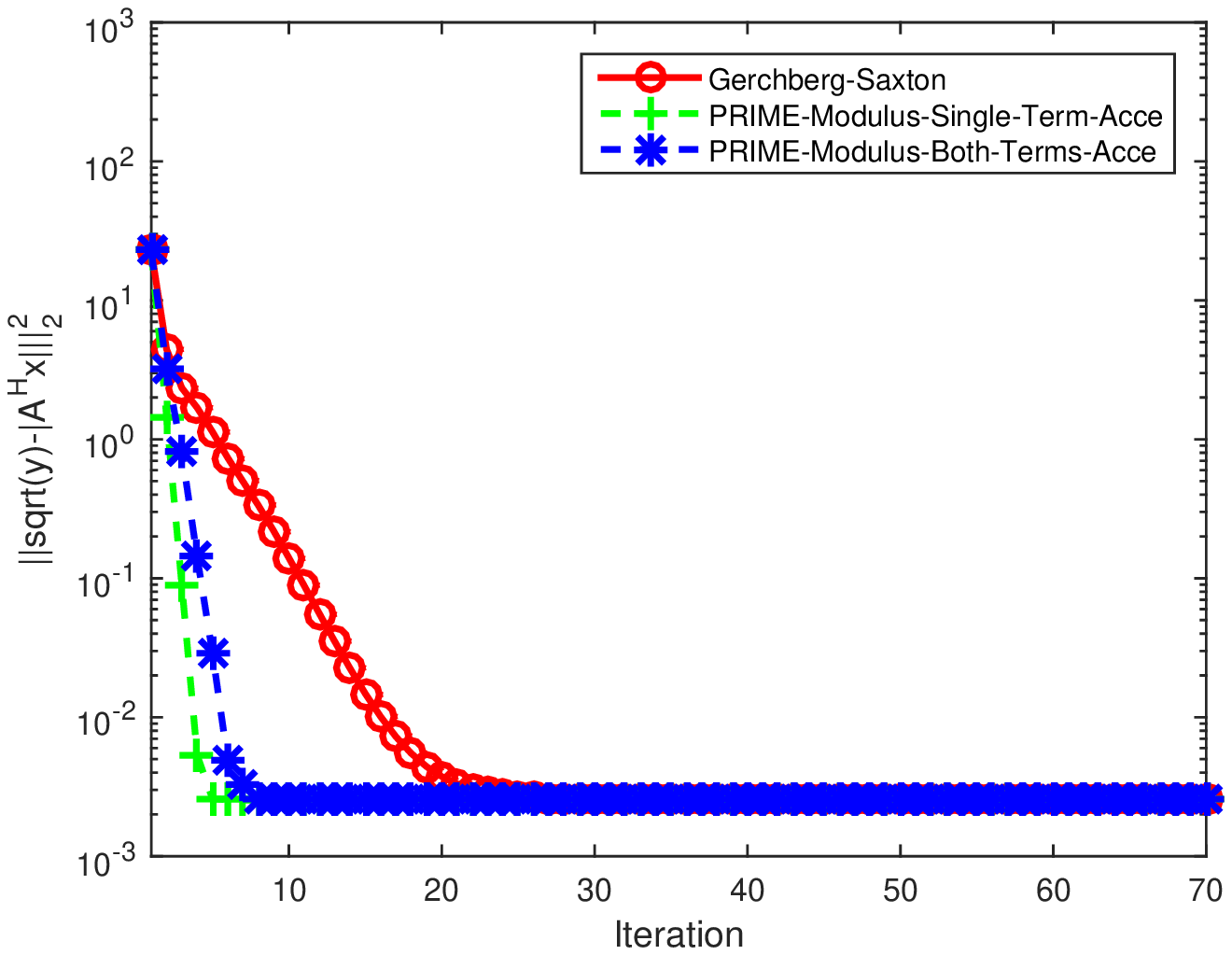}
	\caption{Objective function versus iteration for problem \eqref{eq5} under noisy measurements and random Gaussian matrix setting. $\mathbf{x}\in\mathbb{C}^{10},\mathbf{y}\in\mathbb{R}^{50}$.}
	\label{fig2}
\end{figure}

\begin{figure}[t]
	\centering
	\includegraphics[width=9cm]{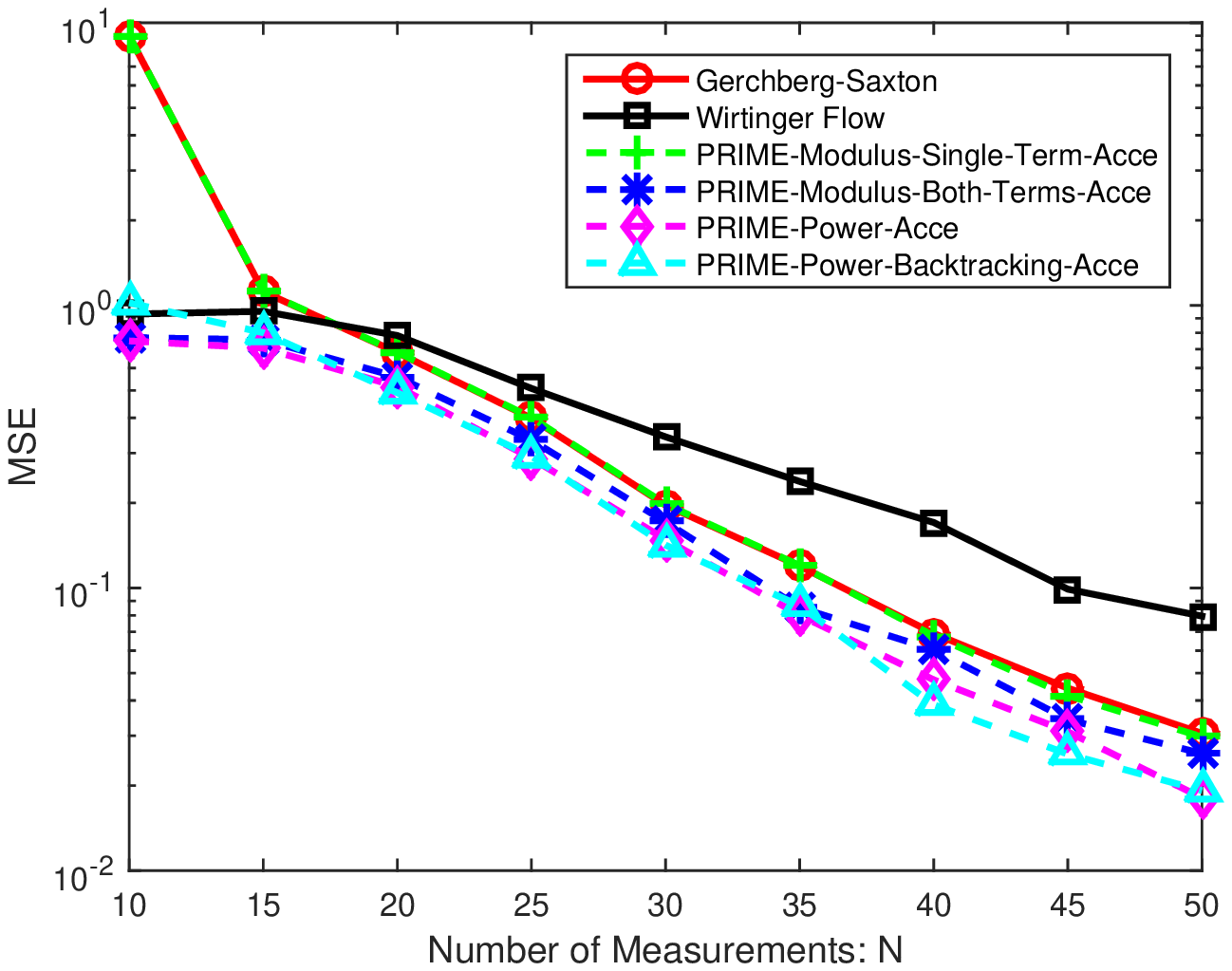}
	\caption{Mean square error (MSE) versus number of clean measurements under random Gaussian matrix setting. $\mathbf{x}\in\mathbb{C}^{10},\mathbf{y}\in\mathbb{R}^N$.}
	\label{fig3}
\end{figure}
\begin{figure}[t]
	\centering
	\includegraphics[width=9cm]{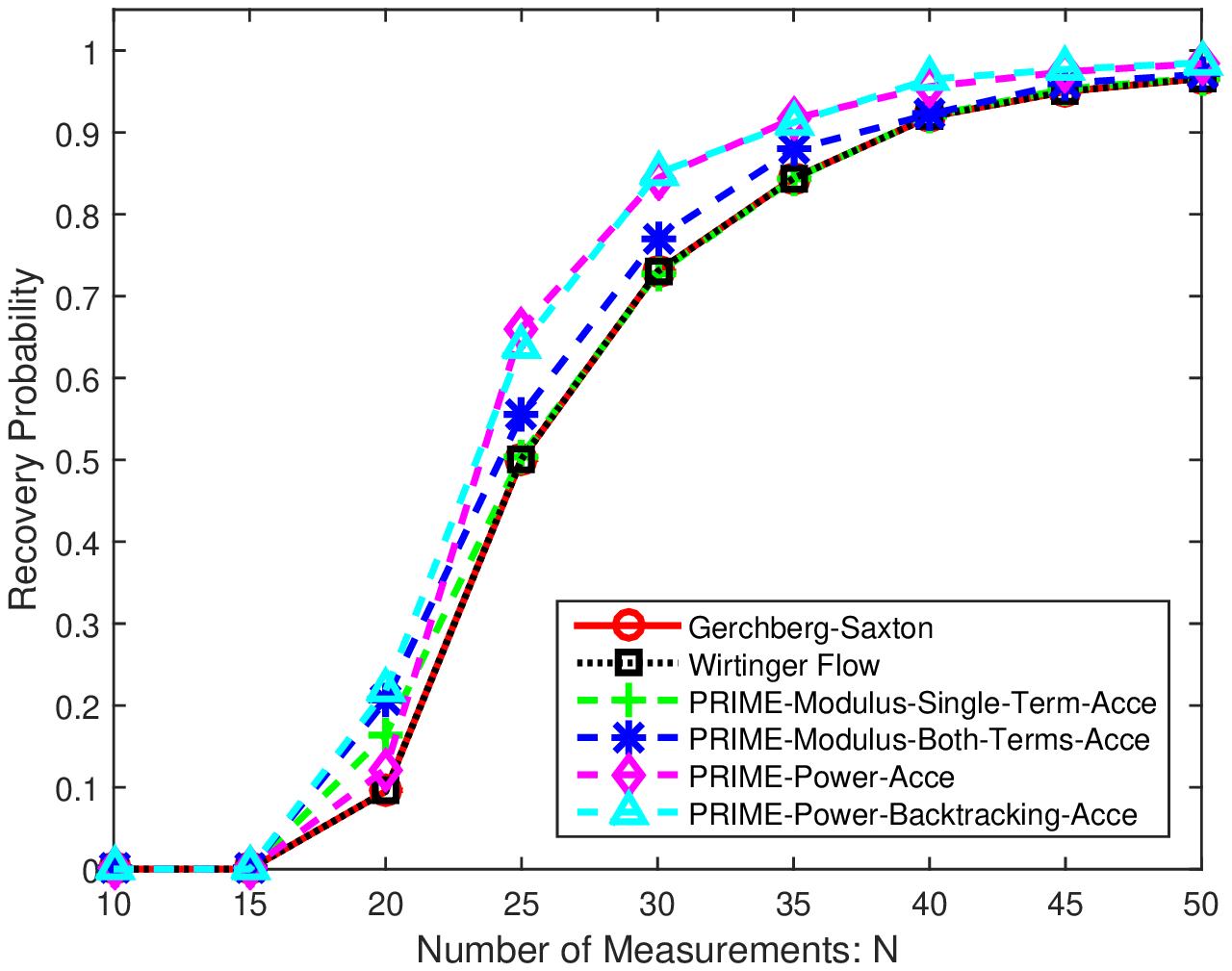}
	\caption{Successful recovery probability versus number of clean measurements under random Gaussian matrix setting. $\mathbf{x}\in\mathbb{C}^{10},\mathbf{y}\in\mathbb{R}^N$.}
	\label{fig4}
\end{figure}

\begin{equation}
	h(\phi)=\left\|\mathbf{x}^{\star}-\mathbf{x}_o\cdot e^{j\phi}\right\|_2^2.
\end{equation}
The derivative of this function $h(\phi)$ with respect to $\phi$ is
\begin{equation}
	\nabla h(\phi)=j\left[\mathbf{x}_o^H\mathbf{x}^{\star}e^{-j\phi}-(\mathbf{x}^{\star})^H\mathbf{x}_oe^{j\phi}\right].
\end{equation}
Setting this derivative to zero, we get
\begin{equation}
	e^{j\phi}=\frac{\mathbf{x}_o^H\mathbf{x}^{\star}}{\left|\mathbf{x}_o^H\mathbf{x}^{\star}\right|}.
\end{equation}
Therefore, we can compute the square error between the solution $\mathbf{x}^{\star}$ returned from our algorithms and the original signal $\mathbf{x}_o$, taking into consideration this global phase shift as $\|\mathbf{x}^{\star}-\mathbf{x}_oe^{j\phi}\|_2^2$. And we plot the mean square error (MSE) between $\mathbf{x}^{\star}$ and $\mathbf{x}_o$ in Figure \ref{fig3}. Besides this, for every single experiment among these $1000$ independent trials, we consider that an algorithm successfully recovers the original signal if the square error is less than $10^{-4}$. And in Figure \ref{fig4}, we plot the probability of successful recovery based on these $1000$ independent trails for all the algorithms.

From Figure \ref{fig4} and Figure \ref{fig3}, we can see that all of our MM-based algorithms have a higher successful recovery possibility and less mean square error than the two benchmark algorithms except PRIME-Modulus-Single-Term-Acce, which can be formulated exactly the same as the Gerchberg-Saxton algorithm when not accelerated. And it agrees with the conjecture in \cite{Bandeira2014} that about $4K$ measurements are needed for a successful recovery with high probability.

As for the phase retrieval problem, more importance should be placed on the successful recovery probability. Therefore, as an example, we only show in Figures \ref{fig1} and \ref{fig2} that our MM-based algorithms converge faster than the benchmark methods for both problems under noisy measurements and the random Gaussian measurement matrix setting.

\begin{figure}[t]
	\centering
	\includegraphics[width=9cm]{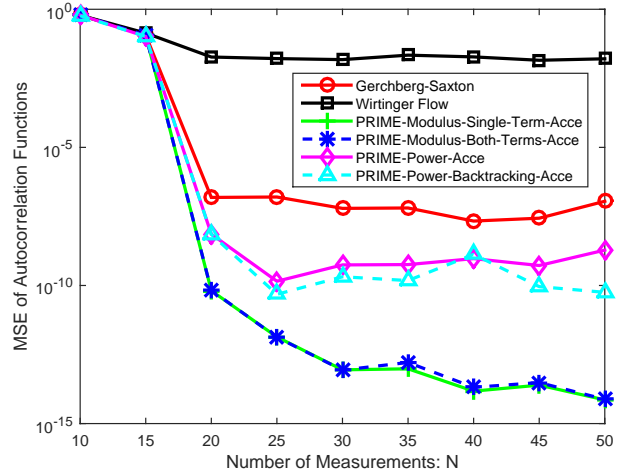}
	\caption{Mean square error (MSE) of autocorrelation functions versus number of clean measurements under DFT matrix setting. $\mathbf{x}\in\mathbb{C}^{10},\mathbf{y}\in\mathbb{R}^N$.}
	\label{fig5}
\end{figure}
\begin{figure}[t]
	\centering
	\includegraphics[width=9cm]{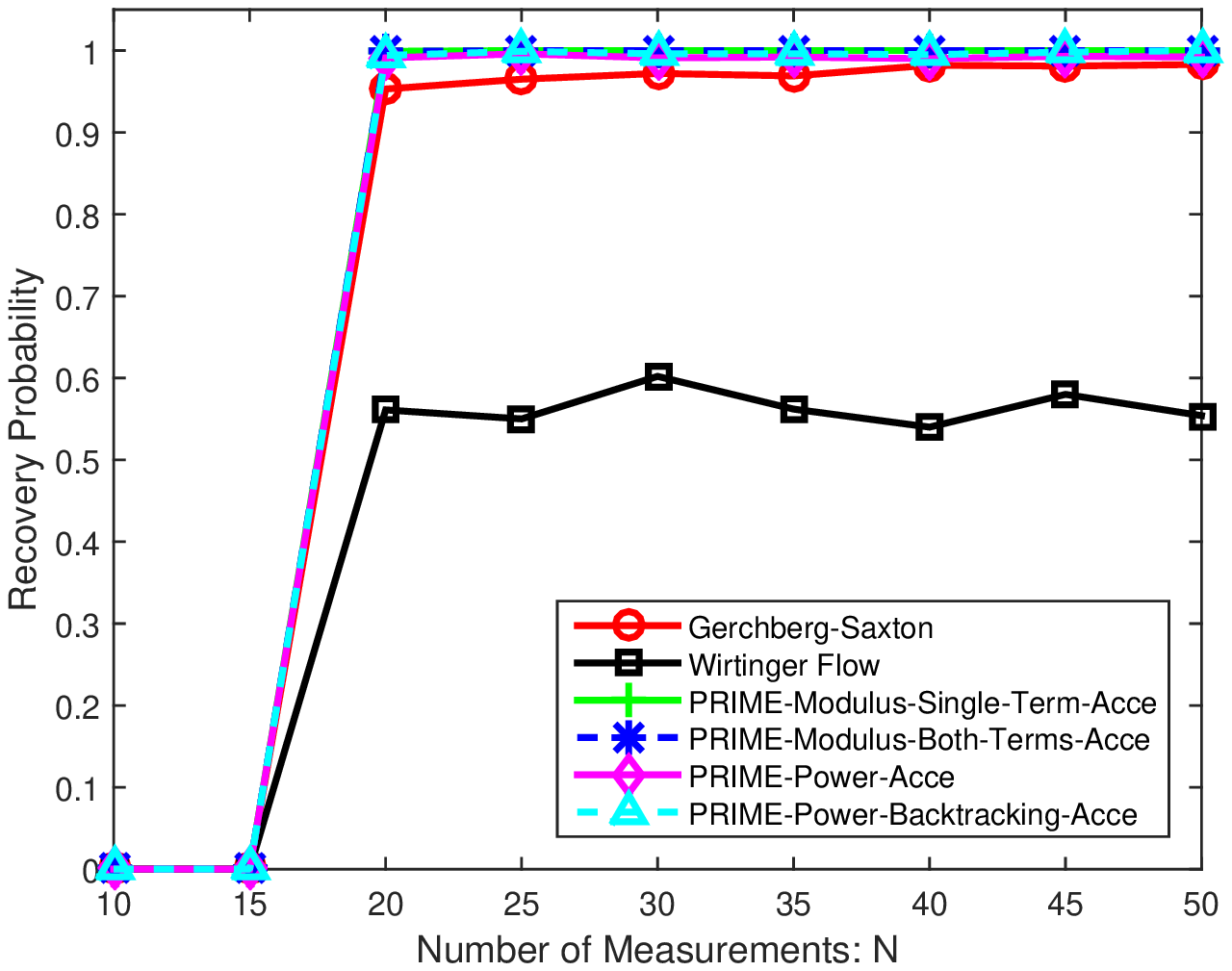}
	\caption{Successful recovery probability of autocorrelation functions versus number of clean measurements under DFT matrix setting. $\mathbf{x}\in\mathbb{C}^{10},\mathbf{y}\in\mathbb{R}^N$.}
	\label{fig6}
\end{figure}

\begin{figure}[t]
	\centering
	\includegraphics[width=9cm]{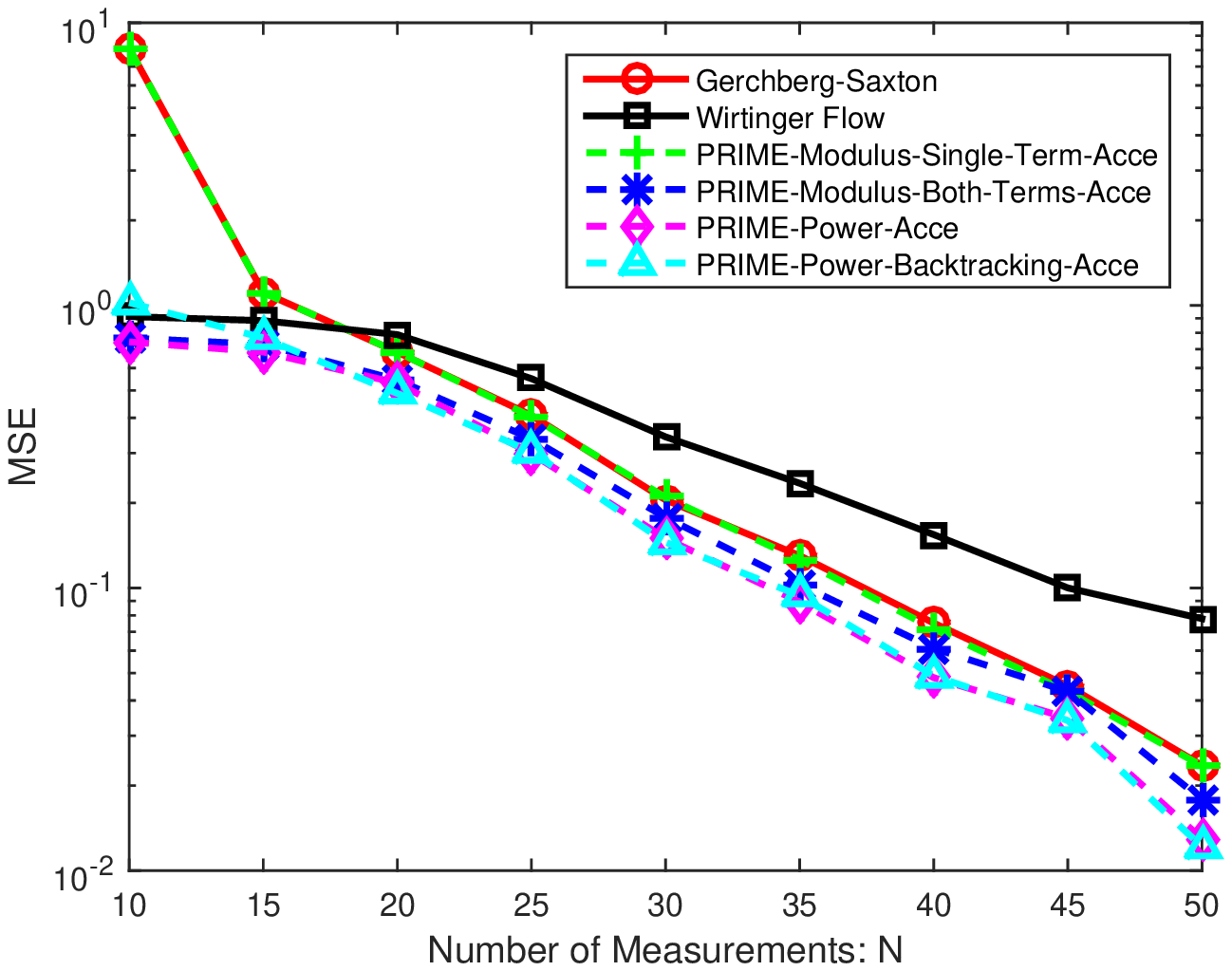}
	\caption{Mean square error (MSE) versus number of noisy measurements under random Gaussian matrix setting. $\mathbf{x}\in\mathbb{C}^{10},\mathbf{y}\in\mathbb{R}^N$.}
	\label{fig7}
\end{figure}
\begin{figure}[t]
	\centering
	\includegraphics[width=9cm]{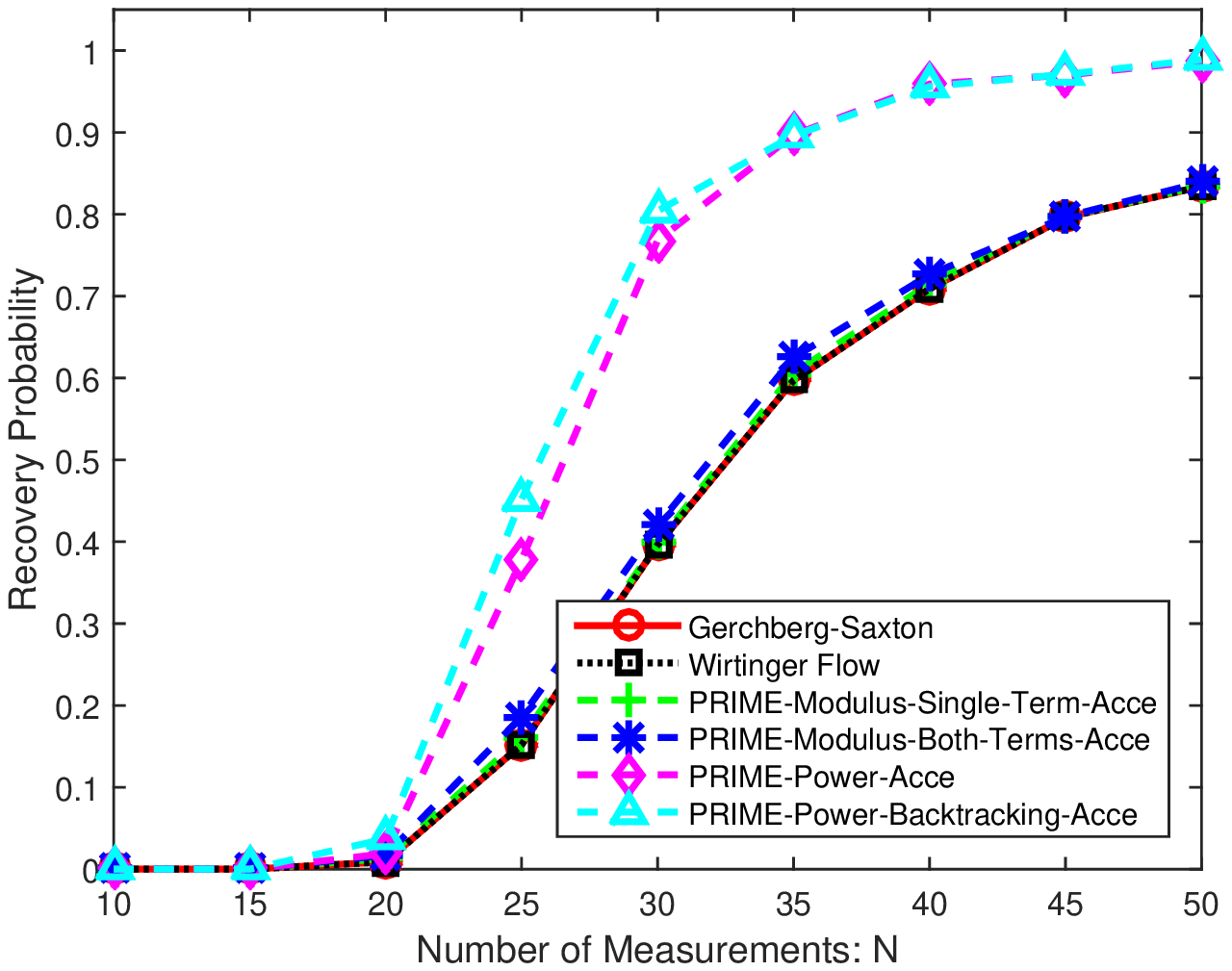}
	\caption{Successful recovery probability versus number of noisy measurements under random Gaussian matrix setting. $\mathbf{x}\in\mathbb{C}^{10},\mathbf{y}\in\mathbb{R}^N$.}
	\label{fig8}
\end{figure}

\begin{figure}[t]
	\centering
	\includegraphics[width=9cm]{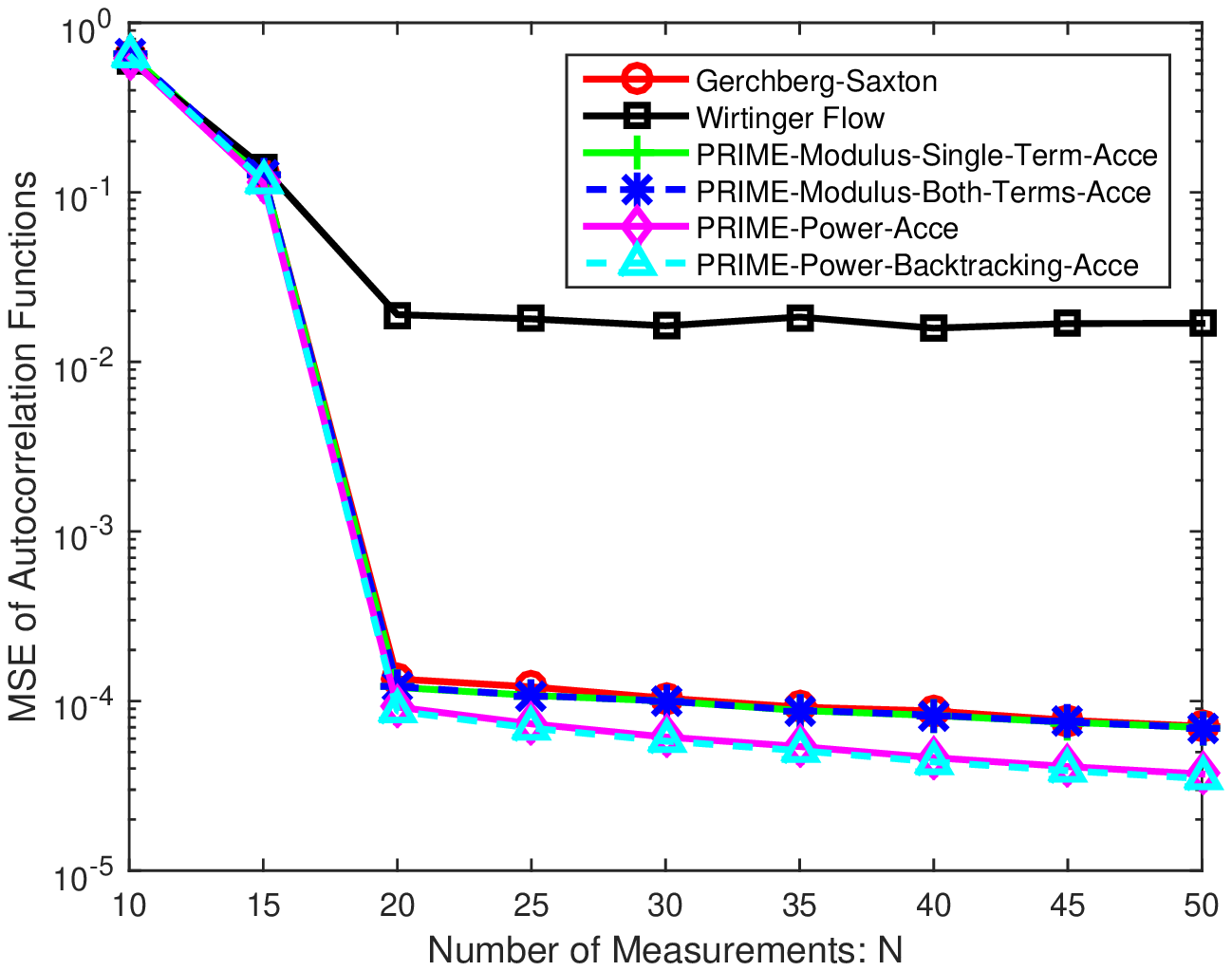}
	\caption{Mean square error (MSE) of autocorrelation functions versus number of noisy measurements under DFT matrix setting. $\mathbf{x}\in\mathbb{C}^{10},\mathbf{y}\in\mathbb{R}^N$.}
	\label{fig9}
\end{figure}
\begin{figure}[t]
	\centering
	\includegraphics[width=9cm]{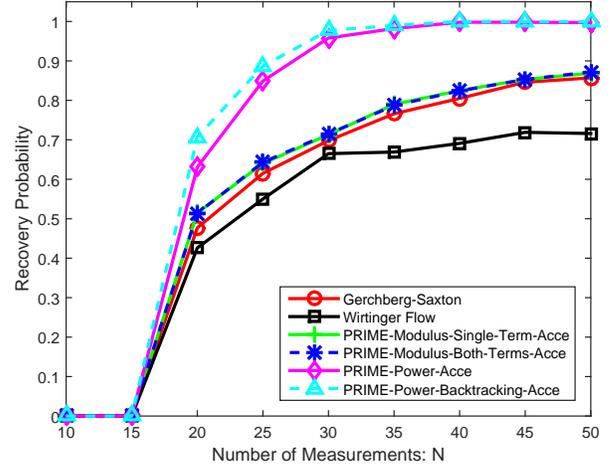}
	\caption{Successful recovery probability of autocorrelation functions versus number of noisy measurements under DFT matrix setting. $\mathbf{x}\in\mathbb{C}^{10},\mathbf{y}\in\mathbb{R}^N$.}
	\label{fig10}
\end{figure}

\subsection{Discrete Fourier Transform Matrix Setting}
In traditional phase retrieval problems, the measurements are the magnitude of the Fourier transform of the signal. Hence, in this subsection, we consider that the measurement matrix $\mathbf{A}$ consists of the first $K$ rows of the $N\times N$ DFT matrix. There are certain advantages to using the DFT properties in our majorization problems. First of all, the leading eigenvalue of the matrix $\mathbf{\Phi}$ is as easy as $\lambda_{\max}(\mathbf{\Phi})=NK$ (proof in Appendix C). And the leading eigenvalue needed in problem \eqref{eq21} also has a simple form now $\lambda_{\max}(\mathbf{AA}^H)=N$ (proof in Appendix A).

Note that there are some differences between the DFT matrix setting and the former random Gaussian matrix setting. The measurement matrix $\mathbf{A}$ is now from the DFT matrix and is not random anymore. The only randomness comes from the original signal $\mathbf{x}_o$. Therefore we need to use different original signals in the $1000$ trials. Another difference is that there are more ambiguities under the DFT matrix setting, unlike under the random Gaussian matrix setting where the global constant phase shift is the only ambiguity. The authors of \cite{Shechtman2015} pointed out that there are always trivial ambiguities and non-trivial ambiguities under the DFT matrix setting for a one dimensional signal. For the trivial ambiguities, any individual or combination of the following three transformations conserve the Fourier magnitude:
\begin{itemize}
	\item[1.]Global constant phase shift: $\mathbf{x}\rightarrow\mathbf{x}\cdot e^{j\phi}$,
	\item[2.]Circular shift: $[\mathbf{x}]_i\rightarrow[\mathbf{x}]_{(i+i_0)\mod{K}}$,
	\item[3.]Conjugate invertion: $[\mathbf{x}]_i\rightarrow\overline{[\mathbf{x}]_{K-i}}$.
\end{itemize}

As for the non-trivial ambiguities, any two signals which have the same autocorrelation function share the same Fourier magnitude. Actually, any two signals within the trivial ambiguities also yield the same autocorrelation function. Therefore under the DFT matrix setting, we can only recover the signal up to the same autocorrelation function without additional information. We use the following autocorrelation function:
\begin{equation}
	[\mathbf{r}]_m=\sum\limits_{i=\max\{1,m+1\}}^{K}[\mathbf{x}]_i\overline{[\mathbf{x}]_{i-m}},\;m=-(K-1),\ldots,K-1.
\end{equation}
And we calculate the autocorrelation function of the original signal $\mathbf{r}_o$ and the autocorrelation function of the solution returned from our algorithms $\mathbf{r}^{\star}$. Later we compute the square error between these two autocorrelation functions $\|\mathbf{r}_o-\mathbf{r}^{\star}\|_2^2$. We also repeat the experiment $1000$ times with different and independent original signals $\mathbf{x}_o$.

In Figure \ref{fig5}, we plot the mean square error of the autocorrelation functions over these $1000$ independent trials, and in Figure \ref{fig6}, we plot the probability for successful recovery based on these $1000$ experiments. In every experiment, an algorithm is considered to successfully recover the signal if the square error $\|\mathbf{r}_o-\mathbf{r}^{\star}\|_2^2$ is less than $10^{-8}$. As shown in Figure \ref{fig6}, all of our MM-based algorithms successfully recover the signal with a higher probability than the benchmark algorithms (although the difference is smaller than in the random Gaussian case). And in Figure \ref{fig5}, our algorithms have less mean square error of the autocorrelation function than those of the benchmark algorithms.

\subsection{Robustness to Noise}
Up to now the experimental results agree with our theoretic analysis that our algorithms outperform the benchmark algorithms under the clean measurements setting. However, in real life the measurements are always corrupted with noise, and usually noise will degrade the performance of an algorithm. Therefore it is necessary to take the noise into consideration. In this subsection, we present the results when the measurements are corrupted with noise.

We add random Gaussian noise to the measurements and then repeat the experiments under the random Gaussian matrix and DFT matrix settings. The expected value of the energy of the noise is $N\times10^{-4}$. The sample means in our experiment are $5.0162\times 10^{-3}$, $4.5129\times 10^{-3}$, $4.0297\times 10^{-3}$, $3.5314\times 10^{-3}$, $2.9928\times 10^{-3}$, $2.5028\times 10^{-3}$, $1.9961\times 10^{-3}$, $1.4964\times 10^{-3}$ and $1.0032\times 10^{-3}$ for $N=50,45,40,35,30,25,20,15$ and $10$, respectively. Results of the mean square error and successful recovery probability are presented in Figures \ref{fig7} and \ref{fig8} for the random Gaussian matrix setting and Figures \ref{fig9} and \ref{fig10} for the DFT matrix setting. All experiments here are also repeated $1000$ times under the same rules as in the clean measurement case. The threshold for the successful recovery is set as $10^{-4}$, which is less than the energy of the noise.

Comparing the successful recovery probabilities (Figures \ref{fig4} and \ref{fig8}, Figures \ref{fig6} and \ref{fig10}), we find that after adding the noise, the gap between the algorithms for problem \eqref{eq2} and the algorithms for problem \eqref{eq5} becomes significantly larger. Under both matrix settings, PRIME-Power-Acce and PRIME-Power-Backtracking-Acce, the two algorithms for problem \eqref{eq2}, have a considerably higher probability of successful recovery. Besides this, the Gaussian noise degrades the performance of the algorithms for problem \eqref{eq5}, namely, the Gerchberg-Saxton algorithm, PRIME-Modulus-Single-Term-Acce, and PRIME-Modulus-Both-Terms-Acce. The successful recovery probabilities of these three algorithms decrease significantly, while the probabilities of the other three algorithms for problem \eqref{eq2} only decrease a little.

As for the mean square error, under the random Gaussian matrix setting, the plots are almost the same in Figures \ref{fig3} and \ref{fig7}. This is because the values of the mean square error are dominated by those experiments with unsuccessful recoveries, which usually have a significantly large square error. And adding small noise cannot change an unsuccessful recovery to a successful one in most cases. Under the DFT matrix setting, the noise concentrates all the values of the square errors between the autocorrelation functions over the $1000$ independent experiments, which leads to the more condensed mean square error plot in Figure \ref{fig9}. As a result, problem \eqref{eq2} is preferable to problem \eqref{eq5} when the measurements are corrupted with Gaussian noise since the former yields the maximum likelihood estimation of the original signal.

\begin{figure}[t]
	\centering
	\includegraphics[width=9cm]{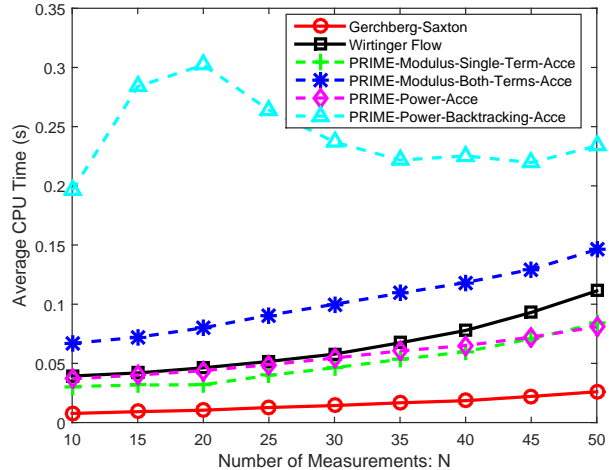}
	\caption{Average CPU Time versus number of noisy measurements under random Gaussian matrix setting. $\mathbf{x}\in\mathbb{C}^{10},\mathbf{y}\in\mathbb{R}^N$.}
	\label{fig11}
\end{figure}

In Figure \ref{fig11}, we plot the average CPU time for all the algorithms over $1000$ Monte Carlo experiments under noisy measurements and random Gaussian measurement matrix setting. For problem \eqref{eq2}, PRIME-Modulus-Single-Term-Acce and PRIME-Modulus-Both-Terms-Acce take slightly more time than the Gerchberg-Saxton algorithm. For problem \eqref{eq5}, PRIME-Power-Acce takes about the same time (less when $N>30$) as the Wirtinger Flow algorithm, but PRIME-Power-Backtracking-Acce takes more time because of the inner loop for the choice of $E$. Although our algorithms require more CPU time, they actually achieve less MSE and higher successful recovery probability as shown in Figures \ref{fig7} and \ref{fig8}. 

\section{Conclusion}
Phase retrieval is of great interest in physics and engineering. Originally the problem was ill-posed due to the loss of phase information. Algorithms based on semidefinite relaxation manage to recover the original signal by solving a convex semidefinite programming problem. But they are not applicable to large scale problems because of the dimension increase in the matrix-lifting procedure. The Wirtinger Flow algorithm recovers the original signal from the modulus square of its linear measurements (problem \eqref{eq2}) using the gradient descent method, but the performance is relatively poor. The classical Gerchberg-Saxton algorithm recovers the original signal from the modulus of its linear measurements (problem \eqref{eq5}) through alternating minimizations by introducing a new variable representing phase information. In this paper we have proposed four efficient algorithms under the majorization-minimization framework. Instead of dealing with the cumbersome phase retrieval problems directly, we have considered different majorization problems which yield a simple closed-form solution via different majorization-minimization techniques. Theoretic analysis as well as experimental results under various settings are also presented in the paper to further validate the efficiency of our algorithms.

\appendices
\section{Proof of $\lambda_{\max}(\mathbf{AA}^H)=N$ for DFT Matrix}
The elements in the DFT measurement matrix $\mathbf{A}\in\mathbb{C}^{K\times N}$ ($K\leq N$) are
\begin{equation}
	A_{ki}=e^{j\frac{2\pi(k-1)(i-1)}{N}},\;k=1,\ldots,K,\text{ and }i=1,\ldots,N.
\end{equation}
Hence the element at the $m$-th row and $n$-th column of the square matrix $\mathbf{AA}^H\in\mathbb{C}^{K\times K}$ is
\begin{equation}
	\begin{aligned}
		&[\mathbf{AA}^H]_{mn}\\
		=&\sum\limits_{k=1}^{N}A_{mk}\overline{A_{nk}}=\sum\limits_{k=1}^{N}e^{j\frac{2\pi(m-1)(k-1)}{N}}e^{-j\frac{2\pi(n-1)(k-1)}{N}}\\
		=&\sum\limits_{k=1}^{N}e^{j\frac{2\pi(m-n)(k-1)}{N}}=\left\{\begin{aligned}
			& N,&& m=n,\\
			& 0,&& \text{otherwise}.
		\end{aligned}\right.
	\end{aligned}
\end{equation}
Thus $\mathbf{AA}^H=N\mathbf{I}_K$. Therefore $\lambda_{\max}(\mathbf{AA}^H)=N$.

\section{Proof of Lemma \ref{lemma2}}
First,
\begin{equation}
	\begin{aligned}
		&\lambda_{\max}(\mathbf{W})\geq\frac{(\mathbf{x}^{(k)})^H}{\|\mathbf{x}^{(k)}\|}\mathbf{W}\frac{\mathbf{x}^{(k)}}{\|\mathbf{x}^{(k)}\|}\\
		=&\|\mathbf{x}^{(k)}\|^2+\frac{1}{D}\sum\limits_{i=1}^{N}\left(y_i-\left|\mathbf{a}_i^H\mathbf{x}^{(k)}\right|^2\right)\frac{\left|\mathbf{a}_i^H\mathbf{x}^{(k)}\right|^2}{\|\mathbf{x}^{(k)}\|^2}.
	\end{aligned}
\end{equation}
If $\mathcal{I}=\emptyset$, $\mathbf{W}$ is positive semidefinite, and it is trivial that $\lambda_{\max}(\mathbf{W})>\lambda_{\min}(\mathbf{W})\geq 0$. When $\mathcal{I}\neq\emptyset$ and $\mathbf{W}$ is not positive semidefinite, defining matrix
\begin{equation}
	\mathbf{Z}:=\sum\limits_{i=1}^{N}\left(y_i-\left|\mathbf{a}_i^H\mathbf{x}^{(k)}\right|^2\right)\mathbf{a}_i\mathbf{a}_i^H,
\end{equation}
then
\begin{equation}
	\frac{1}{D}\lambda_{\min}(\mathbf{Z})\leq\lambda_{\min}(\mathbf{W})<0.
\end{equation}

Therefore, $\lambda_{\max}(\mathbf{W})>|\lambda_{\min}(\mathbf{W})|$ will hold if
\begin{equation}
	\|\mathbf{x}^{(k)}\|^2+\frac{1}{D}\sum\limits_{i=1}^{N}\left(y_i-\left|\mathbf{a}_i^H\mathbf{x}^{(k)}\right|^2\right)\frac{\left|\mathbf{a}_i^H\mathbf{x}^{(k)}\right|^2}{\|\mathbf{x}^{(k)}\|^2}>-\frac{\lambda_{\min}(\mathbf{Z})}{D},
\end{equation}
which is equivalent to
\begin{equation}
	D>-\frac{\lambda_{\min}(\mathbf{Z})}{\|\mathbf{x}^{(k)}\|^2}+\sum\limits_{i=1}^{N}\left(\left|\mathbf{a}_i^H\mathbf{x}^{(k)}\right|^2-y_i\right)\frac{\left|\mathbf{a}_i^H\mathbf{x}^{(k)}\right|^2}{\|\mathbf{x}^{(k)}\|^4}.
\end{equation}
Note that 
\begin{equation}
	-\lambda_{\min}(\mathbf{Z})=\lambda_{\max}(-\mathbf{Z})\leq\sum\limits_{i\in\mathcal{I}}\left(\left|\mathbf{a}_i^H\mathbf{x}^{(k)}\right|^2-y_i\right)\|\mathbf{a}_i\|^2.
\end{equation}
Therefore, $\lambda_{\max}(\mathbf{W})>|\lambda_{\min}(\mathbf{W})|$ will hold if
\begin{equation}
	\begin{aligned}
		D>&\sum\limits_{i\in\mathcal{I}}\left(\left|\mathbf{a}_i^H\mathbf{x}^{(k)}\right|^2-y_i\right)\frac{\|\mathbf{a}_i\|^2}{\|\mathbf{x}^{(k)}\|^2}\\
		&+\sum\limits_{i=1}^{N}\left(\left|\mathbf{a}_i^H\mathbf{x}^{(k)}\right|^2-y_i\right)\frac{\left|\mathbf{a}_i^H\mathbf{x}^{(k)}\right|^2}{\|\mathbf{x}^{(k)}\|^4}.
	\end{aligned}
\end{equation}

\section{Proof of $\lambda_{\max}(\mathbf{\Phi})=NK$ for DFT Matrix}
Recall the definition of the Hermitian matrix
\begin{equation}
	\mathbf{A}_i=\mathbf{a}_i\mathbf{a}_i^H\in\mathbb{C}^{K\times K},\;i=1,\ldots,N,\;K\leq N.
\end{equation}
Hence the element at the $m$-th row and $n$-th column of this square matrix is
\begin{equation}
	\begin{aligned}
		&[\mathbf{A}_i]_{mn}=[\mathbf{a}_i]_m\cdot\overline{[\mathbf{a}_i]_n}=e^{j\frac{2\pi(i-1)(m-1)}{N}}e^{-j\frac{2\pi(i-1)(n-1)}{N}}\\
		=& e^{j\frac{2\pi(i-1)(m-n)}{N}},\;i=1,\ldots,N,\text{ and }m,n=1,\ldots,K.
	\end{aligned}
\end{equation}
So the $((s-1)K+t)$-th element in the vector $\mathrm{vec}(\mathbf{A}_i)$ is
\begin{equation}
	[\mathrm{vec}(\mathbf{A}_i)]_{(s-1)K+t}=[\mathbf{A}_i]_{ts}=e^{j\frac{2\pi(i-1)(t-s)}{N}},\;t,s=1,\ldots,K.
\end{equation}

Also recall the definition of the Hermitian matrix
\begin{equation}
	\mathbf{\Phi}=\sum\limits_{i=1}^{N}\mathrm{vec}(\mathbf{A}_i)\mathrm{vec}(\mathbf{A}_i)^H\in\mathbb{C}^{K^2\times K^2}.
\end{equation}
Thus the element at the $((s_1-1)K+t_1)$-th row and $((s_2-1)K+t_2)$-th column of matrix $\mathbf{\Phi}$ is
\begin{equation}
	\begin{aligned}
		&[\mathbf{\Phi}]_{(s_1-1)K+t_1,(s_2-1)K+t_2}\\
		=&\sum\limits_{i=1}^{N}e^{j\frac{2\pi(i-1)(t_1-s_1)}{N}}e^{-j\frac{2\pi(i-1)(t_2-s_2)}{N}}\\
		=&\sum\limits_{i=1}^{N}e^{j\frac{2\pi(i-1)(t_1-s_1-t_2+s_2)}{N}}\\
		=&\left\{\begin{aligned}
			& N,&&t_1-s_1=t_2-s_2,\\
			& 0,&&\text{otherwise,}
		\end{aligned}\right.\\
		&t_1,t_2,s_1,s_2=1,\ldots,K.
	\end{aligned}
\end{equation}
The summation of all the elements at the $((s_1-1)K+t_1)$-th row of the matrix $\mathbf{\Phi}$ is
\begin{equation}
	[\mathbf{\Phi}\cdot\mathbf{1}]_{(s_1-1)K+t_1}=\sum\limits_{s_2=1}^{K}\sum\limits_{t_2=t_1-s_1+s_2}{}N\leq NK,
\end{equation}
where equality is achieved when $s_1=t_1$.

Note that the matrix $\mathbf{\Phi}$ is a symmetric matrix in which all the elements are real numbers, either $N$ or $0$. And it is also positive semidefinite by the definition. Therefore all the eigenvalues of the matrix $\mathbf{\Phi}$ are nonnegative real numbers. Finally, we adopt the following method to find the leading eigenvalue. For any vector $\mathbf{x}\in\mathbb{C}^{K^2}$,
\begin{equation}
	\begin{aligned}
		&\mathbf{x}^H(NK\mathbf{I}-\mathbf{\Phi})\mathbf{x}\geq\mathbf{x}^H(\mathrm{Diag}(\mathbf{\Phi}\cdot\mathbf{1})-\mathbf{\Phi})\mathbf{x}\\
		=&\sum\limits_{m=1}^{K^2}\overline{x_m}x_m\sum\limits_{n=1}^{K^2}[\mathbf{\Phi}]_{mn}-\sum\limits_{m=1}^{K^2}\sum\limits_{n=1}^{K^2}\overline{x_m}[\mathbf{\Phi}]_{mn}x_n\\
		=&\sum\limits_{m=1}^{K^2}\sum\limits_{n=1}^{K^2}[\mathbf{\Phi}]_{mn}\overline{x_m}(x_m-x_n)\\
		=&\frac{1}{2}\sum\limits_{m=1}^{K^2}\sum\limits_{n=1}^{K^2}[\mathbf{\Phi}]_{mn}\left[\overline{x_m}(x_m-x_n)+\overline{x_n}(x_n-x_m)\right]\\
		=&\frac{1}{2}\sum\limits_{m=1}^{K^2}\sum\limits_{n=1}^{K^2}[\mathbf{\Phi}]_{mn}\left|x_m-x_n\right|^2\geq 0,
	\end{aligned}
\end{equation}
where the third equality comes from the fact that $\mathbf{\Phi}$ is a symmetric real matrix. Therefore,
\begin{equation}
	\lambda_{\max}(\mathbf{\Phi})\leq NK.
\end{equation}
Now we choose $\mathbf{x}=\mathrm{vec}(\mathbf{I}_K)$. Then
\begin{equation}
	\begin{aligned}
		&\frac{\mathbf{x}^H\mathbf{\Phi}\mathbf{x}}{\mathbf{x}^H\mathbf{x}}=\sum\limits_{i=1}^{N}\frac{\mathrm{vec}(\mathbf{I}_K)^H\mathrm{vec}(\mathbf{A}_i)\mathrm{vec}(\mathbf{A}_i)^H\mathrm{vec}(\mathbf{I}_K)}{\mathrm{vec}(\mathbf{I}_K)^H\mathrm{vec}(\mathbf{I}_K)}\\
		=&\sum\limits_{i=1}^{N}\frac{\left(\mathrm{Tr}(\mathbf{A}_i)\right)^2}{\mathrm{Tr(\mathbf{I}_K)}}=N\frac{K^2}{K}=NK.
	\end{aligned}
\end{equation}
Therefore the leading eigenvalue $\lambda_{\max}(\mathbf{\Phi})=NK$.
%
%



%

\bibliographystyle{IEEETran}
\bibliography{../References}


%
%
%




\end{document}